\documentclass[fleqn,usenatbib]{mnras}

\usepackage{newtxtext,newtxmath}
\usepackage[T1]{fontenc}

\DeclareRobustCommand{\VAN}[3]{#2}
\let\VANthebibliography\thebibliography
\def\thebibliography{\DeclareRobustCommand{\VAN}[3]{##3}\VANthebibliography}

\usepackage{graphicx}	% Including figure files
\usepackage{amsmath}	% Advanced maths commands
\usepackage{ulem}
\usepackage{subcaption}
\newcommand{\dtm}{dust-to-metals} % dtm shorthand
\newcommand{\msun}{\,M$_{\odot}$} % msun shorthand

\newcommand{\hMpc}{{\ifmmode{,h^{-1}{\rm Mpc}}\else{$h^{-1}$Mpc}\fi}}
\newcommand{\hkpc}{{\ifmmode{,h^{-1}{\rm kpc}}\else{$h^{-1}$kpc}\fi}}
\newcommand{\hMsun}{{\ifmmode{\,h^{-1}{\rm {M_{\odot}}}}\else{$h^{-1}{\rm{M_{\odot}}}$}\fi}}

\newcommand{\Mstar}{{\ifmmode{,M_{*}}\else{$M_{*}$}\fi}}
\newcommand{\Mhalo}{{\ifmmode{\,M_{\rm halo}}\else{$M_{\rm halo}$}\fi}}
\newcommand{\ltsima}{$\; \buildrel < \over \sim \;$}
\newcommand{\gtsima}{$\; \buildrel > \over \sim \;$}
\newcommand{\lsim}{\lower.5ex\hbox{\ltsima}}
\newcommand{\gsim}{\lower.5ex\hbox{\gtsima}}

\newcommand{\gadget}{\textsc{gadget-4}}
\newcommand{\subfindhbt}{\textsc{subfind-hbt}}
\newcommand{\subfind}{\textsc{subfind}}

\newcommand{\powderday}{\textsc{powderday}}

\usepackage[dvipsnames]{xcolor}
\title{Using dust to constrain dark matter models}

\author[A. J. Ussing et al.]{
Adam J. Ussing,$^{1,2,3}$\thanks{E-mail: aussing@swin.edu.au}
Robert Adriel Mostoghiu Paun,$^{1,2,3}$
Darren Croton,$^{1,2,3}$
Celine Boehm,$^{2,4}$
\newauthor{}
Alan Duffy,$^{1,2}$
and Chris Power,$^{3,5}$
\\
$^{1}$Centre for Astrophysics and Supercomputing, Swinburne University of Technology, PO Box 218, Hawthorn VIC 3122, Australia\\
$^{2}$ARC Centre of Excellence for Dark Matter Particle Physics (CDM)\\
$^{3}$ARC Centre of Excellence for All Sky Astrophysics in 3 Dimensions (ASTRO 3D)\\
$^{4}$School of Physics, The University of Sydney, NSW 2006, Australia\\
$^{5}$International Centre for Radio Astronomy Research (ICRAR), M468, University of Western Australia, 35 Stirling Hwy, Crawley, WA 6009, Australia
}

\date{Accepted 2024 September 27. Received 2024 September 26; in original form 2024 July 11}

\pubyear{2024}

\begin{document}
\label{firstpage}
\pagerange{\pageref{firstpage}--\pageref{lastpage}}
\maketitle

\begin{abstract}
In this paper, we use hydrodynamic zoom-in simulations of Milky Way-type haloes to explore using dust as an observational tracer to discriminate between cold and warm dark matter (WDM) universes. Comparing a cold and 3.5 keV WDM particle model, we tune the efficiency of galaxy formation in our simulations using a variable supernova rate to create Milky Way systems with similar satellite galaxy populations while keeping all other simulation parameters the same. Cold dark matter (CDM), having more substructure, requires a higher supernova efficiency than WDM to achieve the same satellite galaxy number. These different supernova efficiencies create different dust distributions around their host galaxies, which we generate by post-processing the simulation output with the \powderday{} codebase. Analysing the resulting dust in each simulation, we find $\sim$4.5 times more dust in our CDM Milky Way haloes compared with WDM. The distribution of dust out to $R_{200\text{c}}$ is then explored, revealing that the WDM simulations are noticeably less concentrated than their CDM counterparts, although differences in substructure complicate the comparison. Our results indicate that dust is a possible unique probe to test theories of dark matter.

\end{abstract}


%%%%%%%%%%%%%%%%% BODY OF PAPER %%%%%%%%%%%%%%%%%%

\section{Introduction} \label{sec:Introduction}
The leading description of the cosmology of the Universe, known as $\Lambda$ cold dark matter (CDM), combines a model for dark energy, $\Lambda$,, with CDM, some non-interacting massive particles assumed to be non-relativistic during a significant fraction of the evolution of the universe. This description has been remarkably successful in explaining various phenomena in the large-scale Universe, including the angular power spectrum of the cosmic microwave background \citep{aghanim_planck_2020} and baryonic acoustic oscillations \citep{eisenstein_detection_2005}. Nevertheless, despite numerous observations supporting the existence of dark matter \citep{zwicky_rotverschiebung_1933, rubin_rotation_1985, clowe_direct_2006}, a dark matter particle has yet to be found, leaving its true nature as one of the great mysteries in science.

There has been a vast array of experiments designed to detect an imprint of dark matter particles, both through direct and indirect detection methods. Altogether, these experiments have probed an enormous mass range, from $10^{-22}$ eV up to TeV scales (see \cite{klasen_indirect_2015} for a review), though with limited sensitivity. So far, all these experiments have set stringent limits on the dark matter parameter space (mass and interaction rate with ordinary matter or a hypothetical dark sector), except for the DAMA experiment. They claimed the detection of an annual modulation signal compatible with a CDM particle in the late 1990s \citep{bernabei_search_2000}, and maintain this claim as the DAMA/LIBRA experiment \citep{bernabei_first_2008, bernabei_dark_2021}. This signal is in strong tension with the recent findings from the XenonnT \citep{aprile_first_2023}, LZ \citep{aalbers_first_2023} and the plethora of other direct detection experiments, including those using a similar technique \citep{adhikari_strong_2021, coarasa_anais-112_2024}, which is difficult to explain theoretically. There is also a possible hint in favour of dark matter particles from the excess of gamma-rays from the galactic centre found in the FERMI-LAT data \citep{goodenough_possible_2009, hooper_dark_2011}, which is also highly debated (see, e.g. \cite{song_robust_2024} for the latest analysis). If confirmed, this would correspond to a $\sim$30GeV candidate with very suppressed interactions with nuclei to evade the direct detection limits. A dark matter discovery will take more than two highly debated hints (especially given the plethora of negative searches), so the best way to progress is to exploit the wealth of information coming from astrophysical observations. These observations are particularly useful to probe whether dark matter is collisionless \citep[e.g.][]{boehm_constraints_2005, boehm_using_2014, mosbech_probing_2023}, and if so, determine the lower bound of the particle mass \citep[e.g.][]{nadler_constraints_2021}. The abundance of negative direct detection searches combined with the flexibility of fits to astronomical data allows for the possibility of a more exotic dark matter particle.

The physical properties of different candidate dark matter particles manifest as macroscopic differences on astronomical scales, which are often expressed phenomenologically as cold and warm models of dark matter. To combat the difficulties in observing dark matter in the Universe, cosmological simulations have been used to quantify which characteristics of a particular dark matter model best fit the observations. Results from early CDM simulations were able to shed light on several dark matter properties, such as the discovery of a universal halo density profile across a wide range of mass scales, and the abundance of intermediate to high mass haloes \citep{navarro_structure_1996, springel_simulating_2005}. However, these early simulations also resulted in several well-known tensions with observations, for example, the missing satellites problem \citep{klypin_where_1999, moore_dark_1999}, the core-cusp problem \citep{de_blok_core-cusp_2010}, and the too-big-to-fail problem \citep{boylan-kolchin_milky_2012}, see \cite{sales_baryonic_2022} for a recent review. Much research has been dedicated to resolving these tensions by improving the quality of the simulations \citep[e.g.][]{crain_eagle_2015, kaviraj_horizon-agn_2017, pillepich_simulating_2018, hernandez-aguayo_millenniumtng_2022, schaye_flamingo_2023}. These more advanced simulations in recent years bring higher resolutions with better force accuracy and, importantly, the inclusion and advancement of baryonic models.

The missing satellites problem has been the subject of much examination. State-of-the-art $\Lambda$CDM simulations of the 1990s and early 2000s overproduced low-mass dark matter haloes by an order of magnitude compared to expectations based on Milky Way satellite observations of the time. Alternative dark matter models were seen as one possible solution, in particular, warm dark matter (WDM). WDM models introduce a free-steaming scale below which dark matter density fluctuations are erased, thus reducing the number of haloes at high redshift and eventually the number of satellites around Milky Way-like galaxies \citep{bode_halo_2001, avila-reese_formation_2001, boehm_constraints_2005}. The scale at which these low-mass objects are suppressed is set by the mass of the dark matter particle. A heavier WDM particle would look more similar to CDM, whereas a lighter particle produces a much more significant suppression of low-mass halo formation.

The other accepted solution to the missing satellites problem generally invokes baryonic physics to suppress star formation in low mass haloes, often using the feedback of supernova \citep{governato_forming_2007,sawala_apostle_2016}. The energy generated by a supernova event is transferred back into the gas in the simulations, heating the gas. In low-mass haloes with shallow gravitational potential wells, supernova feedback can cause gas outflows with enough energy to eject gas out of the halo entirely, eventually halting star formation. In larger haloes, supernovae still drive outflows, but the potential wells are generally deep enough to retain the gas, which can be accreted back into the galaxy. More efficient supernova models\footnote{Throughout this text, we refer to `supernova efficiency' as the stellar mass fraction converted to supernova, i.e. the $\beta$ parameter in Equation 1 of \cite{springel_cosmological_2003}.}, where stars capable of going supernova are formed at higher rates, are able to suppress star formation in more massive galaxies, though as galaxies get larger AGNs become the dominant form of feedback \citep{croton_many_2006}. Alternatively, reionisation was proposed as a solution to the missing satellites problem \citep{bullock_reionization_2000, benson_effects_2002}, where the photons heat the gas to such a point that low mass haloes can't cool the gas to form galaxies. Importantly, suppression due to reionisation still allows the formation of low-mass galaxies, but preferentially for haloes that form prior to reionisation; thus, the reionisation simulation parameters can be tuned to match satellite observations. 

The tensions between simulations and observations have also been eased in recent years by more complete surveys of the satellite galaxies around the Milky Way \citep[e.g. ][]{mcconnachie_observed_2012,mcconnachie_updated_2020}. These surveys have been able to observe more low luminosity galaxies and galaxy remnants (streams). Combining the updated observations with increasingly precise modelling has led authors such as \cite{sales_baryonic_2022} to conclude that the missing satellites problem can be fully explained within a CDM universe.

However, baryonic solutions to the missing satellites problem come with their own complications. The feedback mechanisms used to reduce low-mass galaxy formation are often conceptually similar but numerically inconsistent between simulations. Supernova feedback alone has an array of implementations across popular codebases, including simply increasing nearby gas temperature \citep{springel_cosmological_2003, governato_forming_2007}, to more complicated kinetic and thermal feedback descriptions \citep{dalla_vecchia_simulating_2008, chaikin_thermal-kinetic_2023}. In reality, there are many forms of feedback, from AGN and supernova to photoheating, which all come with their own implementations within different simulation codes \citep[for a recent review see ][]{vogelsberger_cosmological_2020}. While simulations using a $\Lambda$CDM cosmology can solve the missing satellites problem through high efficiency feedback, the solutions are degenerate with each other and not mutually exclusive of WDM. There is still work to be done to find methods to distinguish between dark matter models.

In this paper, we explore a new method for distinguishing the underlying dark matter species in simulations. It is well known that the missing satellites problem can be solved by implementing either a WDM with reduced feedback efficiency or CDM with increased feedback efficiency. Therefore, we combine each dark matter type with different supernova efficiencies to create simulations with similar Milky Way satellite populations. However, supernovae are also critically important in the creation and transportation of dust \citep{aguirre_intergalactic_1999, todini_dust_2001, bianchi_dust_2007}, and so these different supernova efficiencies are expected to produce different dust properties around the host halo. We explore if these different dust properties can then be used to infer (or at least discriminate) the underlying dark matter model. The techniques we present to analyse the dust provide a proof of concept that can be leveraged in future work with more in-depth analysis of dust in simulations and observations.

This paper is organized as follows. In Section \ref{sec:Simulation}, we discuss the codes used and simulation setup. Section \ref{sec:Results} explains the simulation results with a discussion following in Section \ref{sec:Discussion}. We conclude in Section \ref{sec:Conclusions}. Throughout this paper we assume a flat Planck18 cosmology \citep{aghanim_planck_2020}, where $H_0=66.88$, $\Omega_m=0.321$, $\Omega_{\Lambda}=0.679$, $\Omega_{b}=0.045$, $\sigma_8$=0.840, and $n_s=0.9626$.

\section{Simulations} \label{sec:Simulation}
\subsection{Simulation code setup}\label{sec:code_setup}
For the primary analysis, we perform zoom-in hydrodynamic simulations focusing on Milky Way-like systems. We run cold and warm dark matter, the latter having a mass of 3.5 keV. This is in tension with more stringent observational constraints on the WDM mass \citep{viel_warm_2013, nadler_constraints_2021, nadler_dark_2021, dekker_warm_2022}, but within the limits set by \cite{benito_implications_2020} and  \cite{newton_constraints_2021}. The lighter mass is used to ensure an identifiable difference in the small-scale structure that is wholly attributed to the choice of dark matter.

To generate initial conditions for each cold and WDM simulation, we use \textsc{music}, the Multi-Scale Initial Conditions generator \citep{hahn_multi-scale_2011}. We apply the \cite{eisenstein_baryonic_1998} fitting function to create the input power spectra for the CDM simulations and assume a truncated power spectrum for the WDM simulations, using the \cite{viel_constraining_2005} transfer of the fitting function. The \cite{viel_constraining_2005} parameterisation of the transfer function gives us a WDM half mode mass, $M_{hm} \simeq 2.3\times10^{8}$ \msun{}, at which point we expect significant divergence between cold and warm dark matter simulations. This work is focused on relative differences, so the fitting functions are accurate enough compared to traditional Boltzmann solvers. \textsc{music} disregards any extra thermal velocity the WDM particles may carry in the initial conditions, which can produce extra spurious haloes if not treated correctly \citep{klypin_structure_1993, colin_structure_2008}. Using this generalised approach to generate the WDM power spectrum avoids any assumptions about the nature of the particle itself. The initial conditions for all simulations are created with the same random seed.

We run the parent dark matter only and the subsequent zoom-in simulations with \gadget{} \citep{springel_simulating_2021}, an $N$-body and SPH code with significant improvements to older versions of \textsc{gadget}. \gadget{} makes use of a TreePM algorithm \citep{xu_new_1995, bagla_treepm_2002} for the collisionless particles and a pressure-entropy SPH prescription \citep{springel_cosmological_2002} for the gas particles. The gravitational potential is softened with a cubic spline kernel and becomes Newtonian at distances $r \geq 2.8\epsilon$, where epsilon is the simulation force resolution, and the SPH properties are smoothed over 64 nearest neighbours. The baryonic prescriptions implemented with the public release of \gadget{} are discussed in Section \ref{sec:baryonic_model}. Finally, we use a friends-of-friends algorithm \citep{davis_evolution_1985} and \subfindhbt{} \citep{han_hbt_2018, springel_simulating_2021} for halo and subhalo identification. 

Both parent and zoom-in simulations are run from redshift $z=127$ to $z=0$, with a box side length of 65\hMpc{}, noting that the zoom region is $\sim$0.1\ percent the total volume. The parent simulations have 512$^3$ equal mass dark matter particles, which equates to an approximate mass resolution of $6 \times10^8$\hMsun{} and a force resolution of $10 h^{-1}$kpc. The zoom-in simulations have a dark matter particle resolution of $2.4 \times 10^{6}h^{-1}$\msun{} and a gas particle resolution of $4 \times10^5$\hMsun{}. More details about the simulations can be found in Table \ref{tab:sim_summary}. The parent simulations are used to identify the haloes resembling the Milky Way halo to resimulate.

\begin{table}
    \centering
    \begin{tabular}{c|c|c}
         Parameter & Parent simulation & Zoom simulations\\
         \hline
         Box side length & $65h^{-1}$Mpc & $65h^{-1}$Mpc\\
         Dark matter mass resolution & $5.68 \times$ $10^8h^{-1}$ M$_{\odot}$ & $2.44 \times 10^{6}h^{-1} $M$_{\odot}$\\
         Baryon mass resolution  & NA & $3.99\times 10^{5}h^{-1}$ M$_{\odot}$\\
         Stellar mass resolution & NA & $3.99\times 10^{5}h^{-1}$ M$_{\odot}$\\
         Force resolution & $10h^{-1}$kpc & $500h^{-1}$pc\\

    \end{tabular}
    \caption{Summary of simulation parameters in the parent simulation (left) and the zoom-in re-simulation (right). Box side length is the comoving length of the whole simulation box; the three mass resolutions are the masses of the different particle species, and force resolution refers to the equivalent Plummer softening $\epsilon$, and gravity becomes Newtonian at $2.8\epsilon$.
    }
    \label{tab:sim_summary}
\end{table}

\subsection{Target selection}\label{sec:target_selection}
The halo catalogues produced by \subfindhbt{} \citep{springel_simulating_2021} are used to identify haloes similar in mass to the Milky Way in the parent cold and WDM simulations. We select haloes with a mass $M_{200\text{c}}$ between 1-2$\times 10^{12}h^{-1}$\msun{}, where $M_{200\text{c}}$ is the mass enclosed with a radius $R_{200\text{c}}$ such that the density is 200 times the critical density of the universe, $\rho_\text{c}$. This is consistent with observed Milky Way masses from \cite{bland-hawthorn_galaxy_2016}. We then apply an isolation criterion to these haloes. We exclude any potential resimulation target if there is another halo with a mass greater than 0.5$\times 10^{12}h^{-1}$\msun{} within 8$h^{-1}$Mpc. This ensures that the Milky Way analogue is the largest object in the resimulated region, and that we are not investigating a galaxy undergoing a major merger. We use the same criteria and check both warm and CDM parent simulations to verify the target halo exists in both simulations with similar properties. If there are multiple haloes that fit these criteria, we use the halo closest to the centre of the box.

\subsection{Baryonic model}\label{sec:baryonic_model}
We make use of the baryonic prescriptions that come as part of the public release of \gadget{}. These include a simplified version of the \cite{springel_cosmological_2003} model for star formation and supernova, and the \cite{katz_cosmological_1996} model for radiative cooling. The star formation model is a multiphase model with prescriptions for the separate treatment of hot and cold gas, forming stars from cold gas after reaching a pressure threshold, and adopts a \cite{salpeter_luminosity_1955} initial mass function (IMF). Supernovae are assumed to happen within a single time-step in the simulation, where some fraction of the cold gas designated for star formation is immediately transferred to the hot gas phase. The cooling module treats hydrogen and helium atomic cooling in collisional equilibrium, with the abundances of the ionization states calculated using an external ultraviolet background field, following a modified \cite{haardt_radiative_1996} spectrum \citep{dave_low-redshift_1999}. We adopt this simple baryonic model for straightforward comparisons between our simulations. The baryonic model doesn't include stellar winds or supernova-driven outflows, which is simplistic compared to state-of-the-art models \citep[e.g.][]{crain_eagle_2015, pillepich_simulating_2018, dave_simba_2019}, but more complex models require more complex analysis beyond the scope of this analysis.

We run two simulations for each dark matter type, with different supernovae efficiency parameters, four in total. This is to determine which parameters create the most degenerate solution in the satellite populations when compared with local satellite observations. These are efficiencies of 10 per cent the fiducial value quoted in \cite{springel_cosmological_2003} for a Salpeter IMF, and 5 per cent, knowing that a reduced supernovae efficiency is required in a WDM simulation to produce similar numbers of low mass galaxies as a CDM simulation. As highlighted in Section \ref{sec:Introduction}, the efficiencies here refer to the $\beta$ parameter used in \cite{springel_cosmological_2003}, which is the mass fraction of new stars that become supernova within one simulation time-step.

\subsection{Dust production}\label{sec:dust_production}
To produce dust throughout the galaxy in our simulations, we use the post-processing tool \powderday{} \citep{narayanan_powderday_2021}, which is built on \textsc{yt} \citep{turk_yt_2011}, \textsc{hyperion} \citep{robitaille_hyperion_2011}, and \textsc{fsps} \citep{conroy_propagation_2010}. \powderday{} is designed to provide observational properties of galaxies in cosmological simulations by assigning dust masses to the simulation gas particles according to one of several prescriptions based on observational correlations. These include a dust-to-metals ratio, a gas-to-dust ratio from \cite{remy-ruyer_gas--dust_2014}, and a `best-fit' metallicity dependant gas-to-dust ratio from \cite{li_dust--gas_2019}.

All models are explored in this work. However, for clarity and simplicity, we focus on our results using a dust-to-metals ratio with a value of 0.4 \citep{dwek_evolution_1998, yajima_observational_2015}. In this case, the dust mass is calculated by multiplying the gas mass by metallicity by 0.4, where the default metal yield in \gadget{} is 0.02. A short comparison between dust production methods is presented in Appendix \ref{sec:app_dust_comp}.

\subsection{Gini coefficients}\label{sec:gini_intro}
The Gini coefficient is used to characterize the spatial distribution of dust produced in each simulation. Gini coefficients typically measure uniformity in a sample. When used in processes such as galaxy imaging, the scores typically represent photon flux across pixels in a CCD, where a value of 0 would mean perfect equality in the flux across all pixels, and 1 indicates all flux exists in a single pixel \citep{abraham_new_2003, lotz_new_2004, zamojski_deep_2007}.

In our analysis, we use the Gini coefficient to describe the concentration of dust in the halo. We create a 2D projection map of the dust mass in the halo, with a bin width of $2.5h^{-1}$ckpc, and use the total mass in each bin to calculate the Gini score, $G$:

\begin{equation}\label{eq:gini}
    G = \frac{\sum\limits_{i=1}^n \sum\limits_{j=1}^n |x_i - x_j |}{2n^2\overline{x}},
\end{equation}

where $x_n$ is the value in the $n$th cell, and $\overline{x}$ is the average over $n$ cells. The dust mass is calculated out to a radius of $500h^{-1}$ckpc, which equates roughly twice $R_{200\text{c}}$ at $z=0$; however, bins outside of $R_{200\text{c}}$ or within a circle with a 15ckpc radius at the centre of the halo are masked to exclude them from the Gini calculation. The former avoids counting bins beyond the halo with no dust, which would lower the Gini score, and the latter avoids the highly concentrated central galaxy, which saturates any signal. To avoid projection effects, where satellites might lie directly behind or in front of the central galaxy, we rotate the dust projection maps around each axis. We use 25 viewing angles between 0 and 2$\pi$ in the $x-y$, $x-z$, and $y-z$ axes, then take the median of the 75 individual Gini score calculations. The masking is applied in every projection at all viewing angles. This method is explored further in Appendix \ref{sec:app_halo_mask}.

\section{Results} \label{sec:Results}
\subsection{Milky Way reproduction}\label{sec:mw_reproduction_results}
\begin{table*}
    \centering
    \begin{tabular}{c|c|c|c|c|c|c}
         Simulation & R$_{200\text{c}}$ (kpc) & M$_{200\text{c}}$ ($10^{12}$M$_{\odot}$) & M$_{\text{DM}}$ ($10^{12}$M$_{\odot}$) & M$_{\text{gas}}$ ($10^{10}$M$_{\odot}$) & M$_{\text{stars}}$ ($10^{10}$M$_{\odot}$)& SFR (M$_{\odot}$yr$^{-1}$)\\
         \hline
         CDM - SN 5\%  & 261.8 & 1.9 & 1.6 & 12.9 & 12.0 & 2.5 \\
         
         CDM - SN 10\% & 261.6 & 1.9 & 1.6 & 13.8 & 10.8 & 1.6 \\
         
         WDM - SN 5\%  & 254.1 & 1.7 & 1.5 & 11.0 & 11.6 & 1.4 \\
         
         WDM - SN 10\% & 254.0 & 1.7 & 1.5 & 11.1 & 11.4 & 1.6 \\
         
         Milky Way & 169-284 & 0.55-2.62 & - & 1-5 & 4-7 & 1-3 \\
    \end{tabular}
    \caption{Summary halo information for the Milky Way analogue in each simulation. All units are physical with $h$ factored in. $M_{200\text{c}}$ refers to the total mass within the radius $R_{200\text{c}}$, $M_{\text{DM}}$, $M_{\text{gas}}$, and $M_{\text{stars}}$ refer to the dark matter, gas, and stellar masses, respectively, within the same halo radius. SFR is the star formation rate within the Milky Way analogue. Ranges for the observed Milky Way values are taken from \protect\cite{bland-hawthorn_galaxy_2016}.}
    \label{tab:sim_mass_summary}
\end{table*}    

Before analysing the simulations, we convert their comoving units to physical units and remove the $h$ dependence to compare with each other and later observational data. We then verify that our simulations all produce similar halo properties at $z=0$. Each pair of dark matter simulations produces very similar Milky Way-like galaxies, with per cent level differences in radius, total mass, and dark matter mass, which are summarized in Table \ref{tab:sim_mass_summary}. There are minor differences in the gas and stellar mass, which is to be expected from the differing supernova efficiencies. Comparing the two dark matter models against each other, we see that CDM produces slightly larger haloes on average than WDM, approximately 10 per cent larger in total mass and  3 per cent larger radii. Minor differences between the haloes are to be expected, as the different dark matter models natively produce different formation histories, and the supernova efficiencies give each galaxy a unique evolution. However, these differences between the simulations are within the expected numerical variance. 

Table \ref{tab:sim_mass_summary} also shows the observed values for selected Milky Way properties, taken from \cite{bland-hawthorn_galaxy_2016}, against which we can compare the four simulations. Our resimulation target selection process ensures that we recover a total halo mass in the range of 1-2$\times10^{12}M_{\odot}$. Our stellar masses are generally consistent, though a little larger than the observations, which tend to fall between 4 and 7$\times 10^{10}$\msun{} \citep[see also][]{kafle_shoulders_2014, licquia_improved_2015, mcmillan_mass_2017}. Star formation rates also agree with the observations, which are between 1 and 3 M$_{\odot}$ yr$^{-1}$ \citep{chomiuk_toward_2011, licquia_improved_2015, elia_star_2022}. See \cite{bland-hawthorn_galaxy_2016} for a review of the Milky Way properties.

\subsection{Calibrating the simulations}\label{sec:tuning_results}
\begin{figure}
    \centering
    \includegraphics[width=\columnwidth]{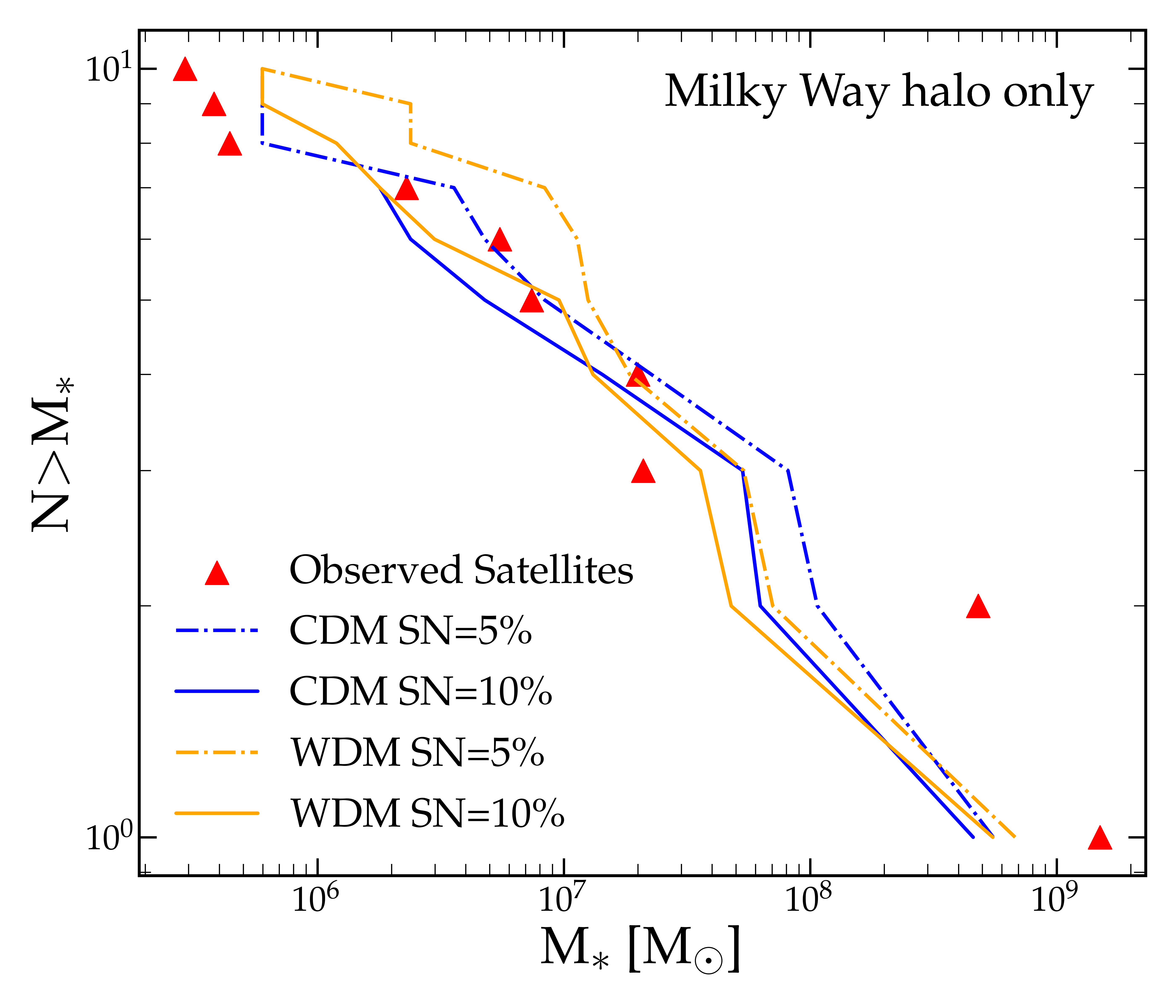}
    \caption{The cumulative satellite galaxy population around each Milky Way analogue in our warm and cold dark matter simulations, with two levels of supernova efficiencies. Blue lines indicate CDM, and orange for WDM; dashed–dotted lines indicate 5 per cent supernova efficiency and solid lines 10 per cent. The red triangles mark the ten largest observed Milky Way satellites; their masses are taken from the compilation by \protect\cite{lovell_matching_2023}.}
    \label{fig:subhalo_central_pop}
\end{figure}

The most significant point of difference between the dark matter models is the satellite galaxies, which we now compare in our simulations. We plot the cumulative satellite mass function for the Milky Way analogues in Fig. \ref{fig:subhalo_central_pop}, with the ten brightest Milky Way satellites overplotted in red triangles. We use observed satellite masses from Table 1 of \cite{lovell_matching_2023} and references therein. For this work, we define a satellite galaxy in the simulation as a subhalo with at least one stellar particle and at least 100 total particles, corresponding to a minimum stellar mass of $6\times10^{5}$\msun{} and a minimum total mass of $6\times10^{8}$\msun{}. The minimum total particle number limit is implemented to try to avoid artificial haloes known to plague WDM simulations \citep{wang_discreteness_2007, power_spurious_2016}. In this setup, these artificial haloes will dominate the mass function at $M_{\text{lim}}\simeq2\times10^8$\msun{} \citep{wang_discreteness_2007, lovell_properties_2014}, which is below our satellite mass resolution limit. While we include anything with a singular star particle, most satellites contain $\gtrsim$10 star particles and generally several hundred gas and dark matter particles. Contrary to expectations, the WDM simulation with 5 per cent supernova efficiency produces the most satellite galaxies of all the simulations. We expect this is a unique case where the satellites have merged at different times rather than a trend in the simulation. We discuss this further in Section \ref{sec:MW_sats}. All the simulations show good agreement with the observed number of satellites, but with so few, it's difficult to distinguish the effect of the combination of dark matter models and feedback efficiencies.

\begin{figure}
    \centering
    \includegraphics[width=\columnwidth]{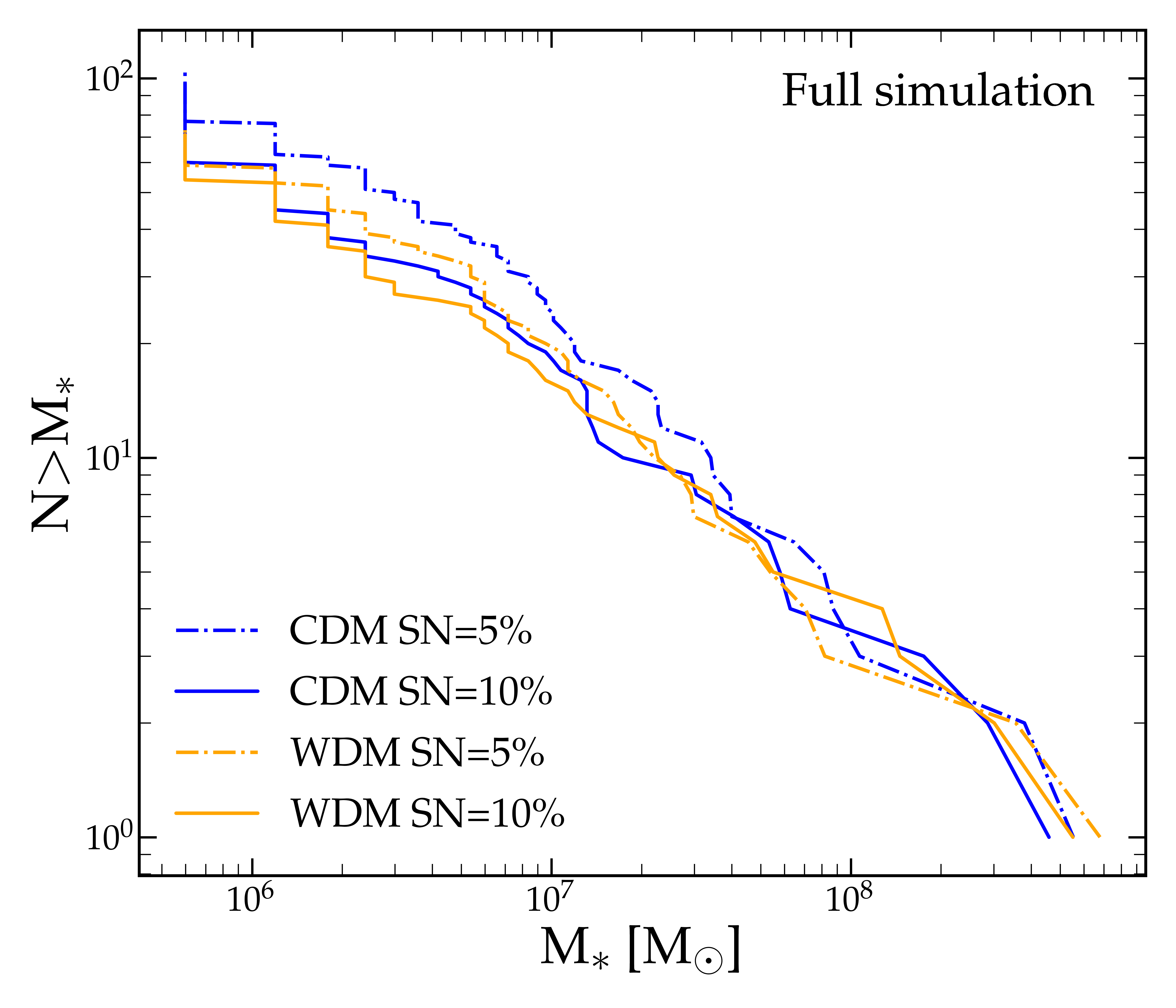}
    \caption{The complete cumulative satellite galaxy population (not just the Milky Way analogue) within each simulation box. We see general convergence between the models in the high-mass satellites, and a divergence between the different dark matter models at the low-mass end, where the impact of supernovae efficiency also becomes more evident. The most degenerate solutions are a CDM simulation with 10 per cent supernova and WDM with 5 per cent. }  
    \label{fig:subhalo_population}
\end{figure}

To better highlight the satellite population degeneracy between simulations, in Fig. \ref{fig:subhalo_population}, we present the entire population of satellite galaxies. This figure shows a similar cumulative mass function, but now for all satellites in the box rather than only those around the Milky Way analogue. In order to be classified as a satellite in this context, the FOF group must contain no low-resolution dark matter particles. Additionally, the subhalo must contain at least one star particle and must not be rank 0, the \subfind{} convention for identifying the host halo. We see the expected divergence at lower masses, with both WDM simulations producing fewer satellites overall than their CDM counterparts. The different supernova efficiencies also show the expected behaviour, with higher rates producing fewer satellites. We find a relatively consistent match between the CDM simulation with a 10 per cent supernova efficiency rate and the WDM simulation with a 5 per cent rate. We designate these two simulations as our fiducial models for our analysis below. 

\subsection{Dust comparison across the simulations}\label{sec:dust_sim_results}
\begin{table}
    \centering
    \begin{tabular}{c|c|c|c}
         &  Dust Mass & Dust Mass & Dust Mass\\ 
         Simulation & R$_{200\text{c}}$ ($10^7$M$_{\odot}$) & 15kpc ($10^7$M$_{\odot}$) & 15kpc (\%)\\
         \hline 
         CDM - SN 5\%  & 1.44  & 1.22  & 85\% \\
         CDM - SN 10\% & 4.56  & 3.98  & 87\% \\
         WDM - SN 5\%  & 1.05  & 0.79  & 75\% \\
         WDM - SN 10\% & 2.52  & 2.36  & 93\% \\
    \end{tabular}
    \caption{The total dust mass produced in our four simulations (see Table \ref{tab:sim_summary}), measured both within the halo radius and the inner 15kpc around the central Milky Way analogue galaxy. The majority of the dust is concentrated within the central galaxy of the halo.} 
    \label{tab:dust_masses}
\end{table}

After verifying that two of our models create degenerate satellite galaxy populations, both with each other and observed Milky Way satellites, we now study the dust in each simulation to determine whether it can be used as an observable to distinguish between cold and warm dark matter. 

Table \ref{tab:dust_masses} shows the total dust mass across all simulations, within $R_{200\text{c}}$ and the central 15kpc, using the \dtm{} prescription described in Section \ref{sec:dust_production}. Crucially, the difference between the fiducial cold (10 per cent) and warm (5 per cent) dark matter models (the two best matching configurations) is significant, with $>4$ times the total amount of dust in the CDM simulation. An increase in supernovae efficiency increases the amount of dust in both dark matter models, though less so for WDM. The lower supernova efficiency effectively locks away more gas in stars, reducing the availability of gas for star-forming episodes that drive increases in metallicity. The concentration, or the fraction of dust within 15kpc of the halo centre, ranges from 75 per cent in the fiducial WDM simulation to up to 87 per cent in the fiducial CDM simulation. This presents the possibility that direct observations of the dust inside a galaxy can be used as a metric to differentiate the nature of the dark matter species, in terms of aggregated dust masses.

\begin{figure*}
    \begin{subfigure}[t]{0.32\textwidth}
        \includegraphics[width=\linewidth]{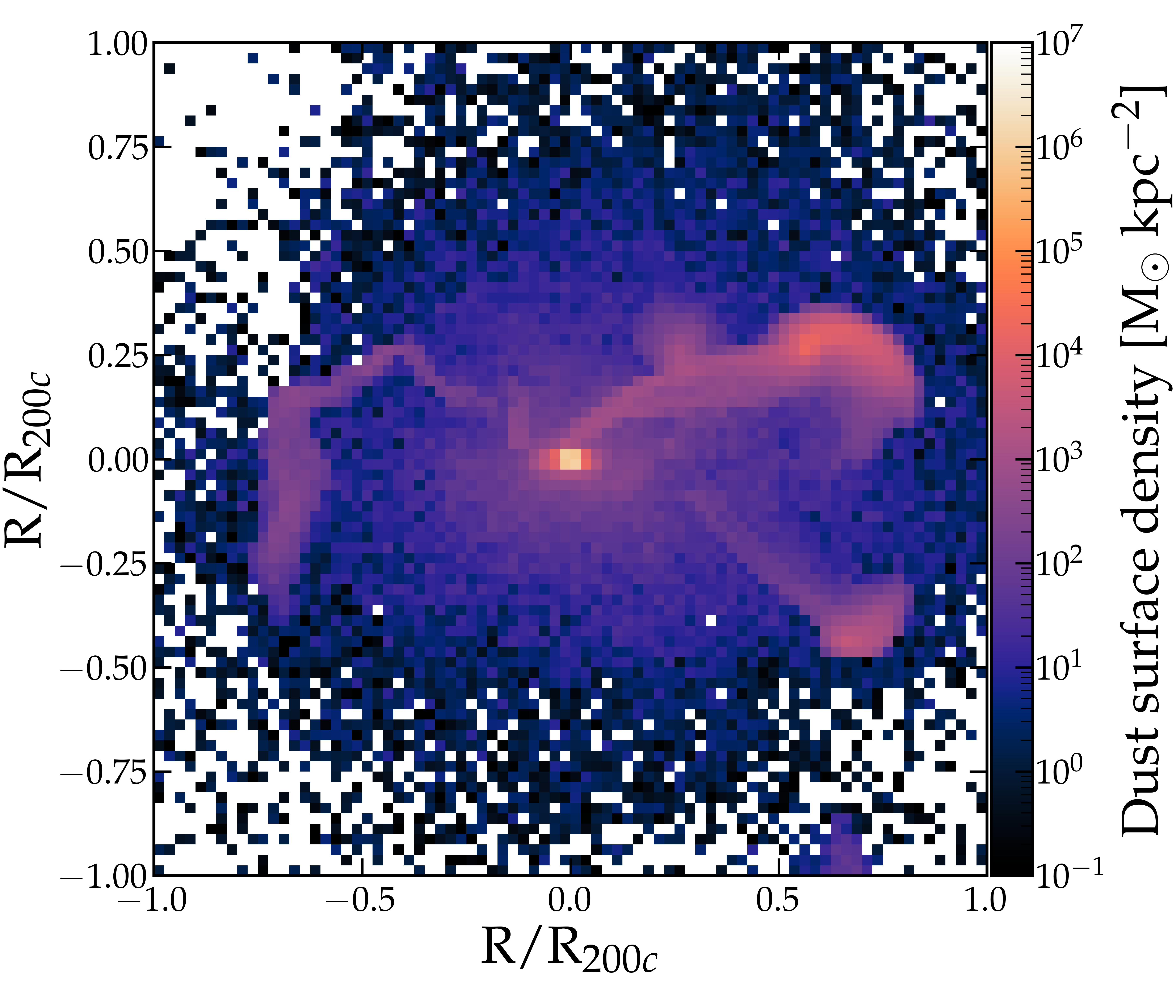}
        \caption{Cold dark matter - 10\% Supernova efficiency}
        \label{fig:cdm_dust_hist}
    \end{subfigure}
    \hfill
    \begin{subfigure}[t]{0.32\textwidth}
        \includegraphics[width=\linewidth]{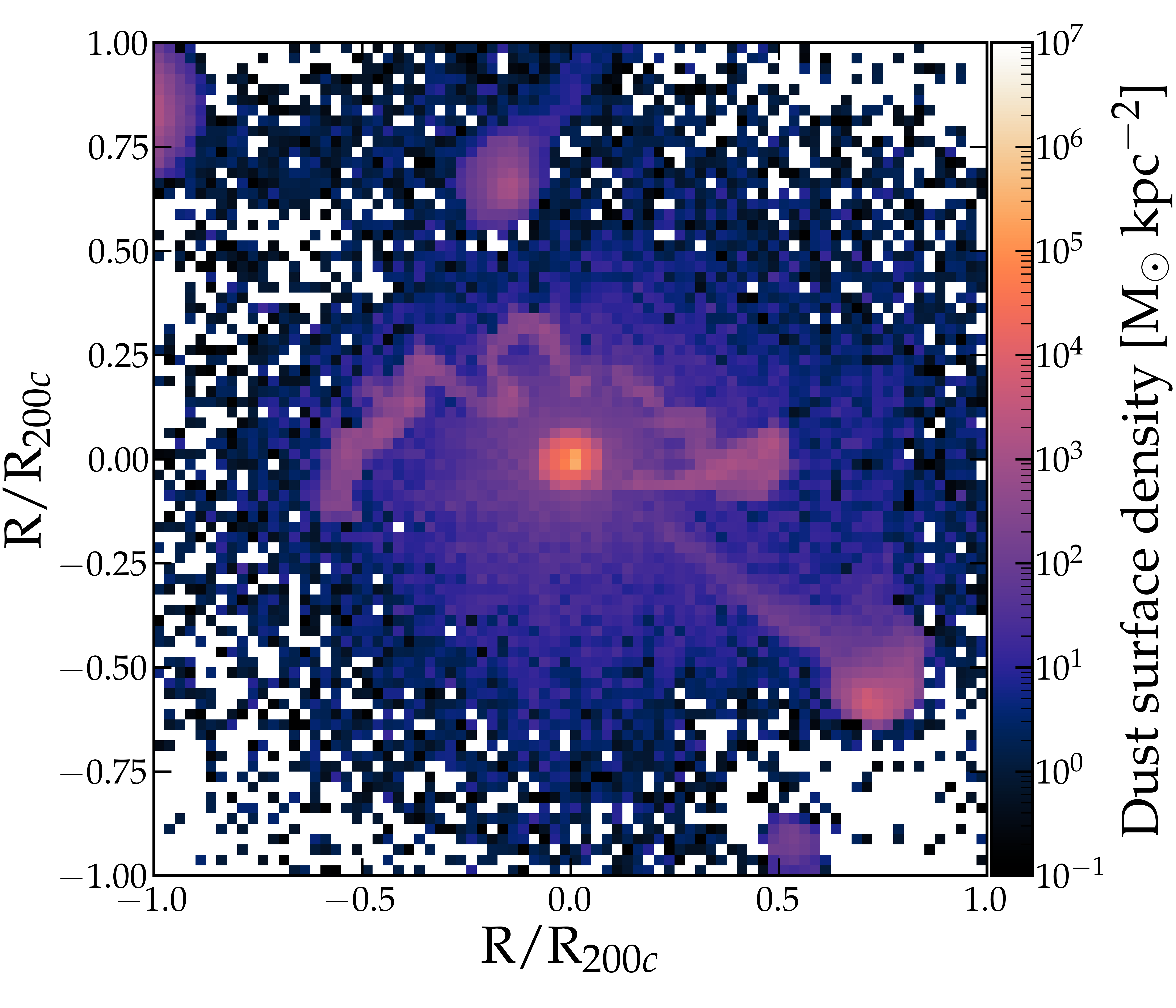}
        \caption{Warm dark matter - 5\% Supernova efficiency}
        \label{fig:wdm_dust_hist}
    \end{subfigure}
    \hfill
    \begin{subfigure}[t]{0.32\textwidth}
        \includegraphics[width=\linewidth]{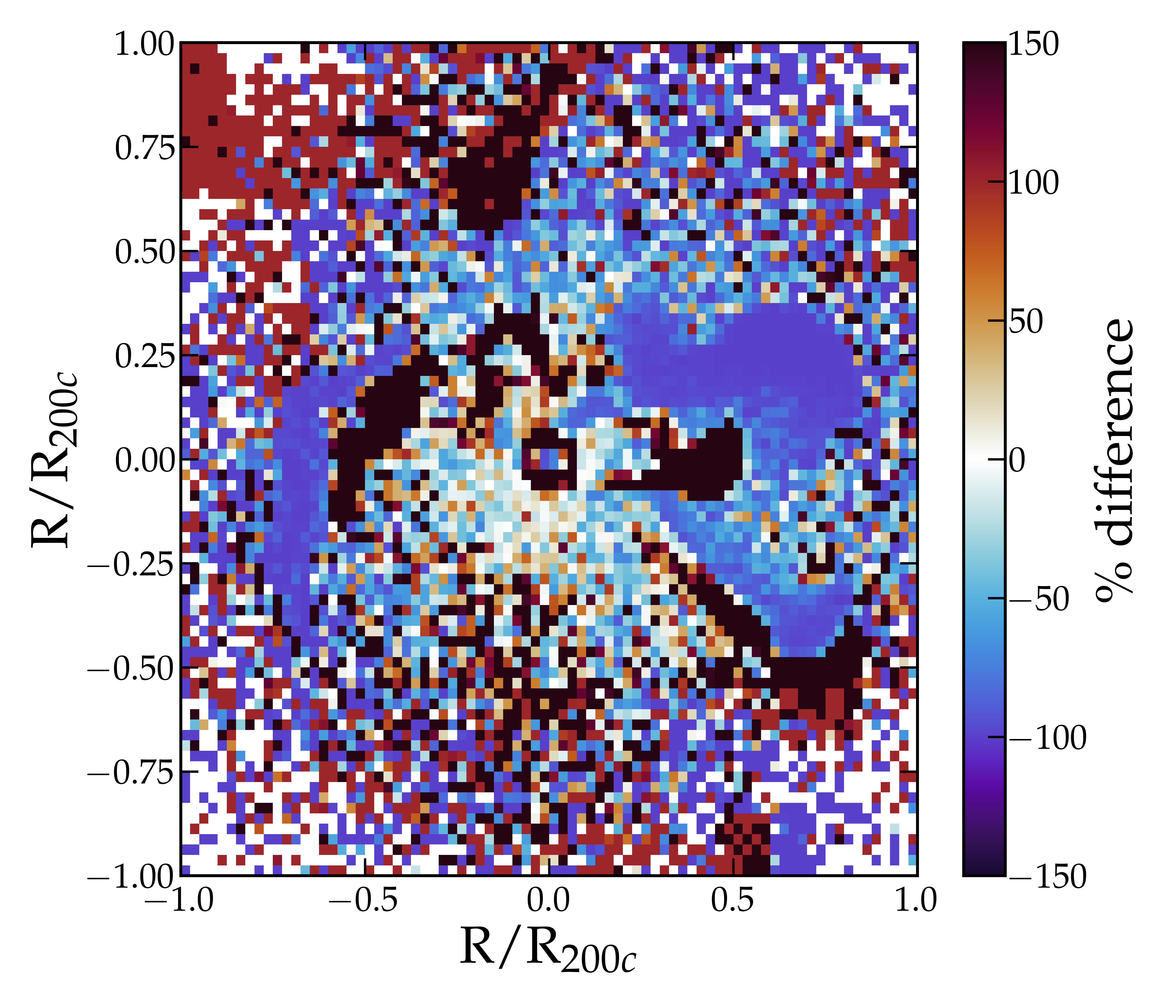}
        \caption{Residuals}
        \label{fig:dust_diff}
    \end{subfigure}
    \caption{\textit{Left and Middle:} Dust map projection plots around our Milky Way analogue galaxies, showing CDM with a 10 per cent supernovae efficiency on the left, and WDM with 5 per cent supernovae efficiency in the middle. The projection extends out to $R_{200\text{c}}$. \textit{Right:} The percentage difference of the dust mass between the two simulations. High values in red show more dust in the WDM simulation and low values in blue show more dust in CDM.
    }
    \label{fig:dust_hists_all}
\end{figure*}

To explore dust outside of the galaxy (but inside the halo), Figs \ref{fig:cdm_dust_hist} and \ref{fig:wdm_dust_hist} plot the 2D projections of its distribution out to the halo radius for our fiducial cold and warm dark matter simulations. Beyond the central galaxy, the remaining dust tends to be concentrated in the satellites, with lower densities in the diffuse gas. The differential map between the two simulations is shown in Fig. \ref{fig:dust_diff}. The colour bar shows the percentage difference between the two, with blue indicating higher dust mass in the CDM simulation and red indicating higher dust mass in the WDM simulation. This comparison shows the same behaviour as that of the total mass, that generally, there is more dust in the CDM simulation spread throughout the halo. Besides this general interpretation, the differential map has several key features. The most obvious is that the difference in satellite positions creates stark differences in the differential dust map, even if their positions are only marginally offset. The other interesting feature is the difference in dust distribution around the central galaxy. The CDM simulation has a much more compact dust disc than the WDM simulation, which is primarily driven by the difference in supernova efficiencies. When we switch the efficiencies, i.e. CDM with a 5 per cent supernova efficiency and WDM with a 10 per cent efficiency, we see the opposite trend, where the CDM has a much more extended dust disc. This switch does not significantly affect the dust outside of the central galaxy.

\begin{figure}
    \centering
    \includegraphics[width=\columnwidth]{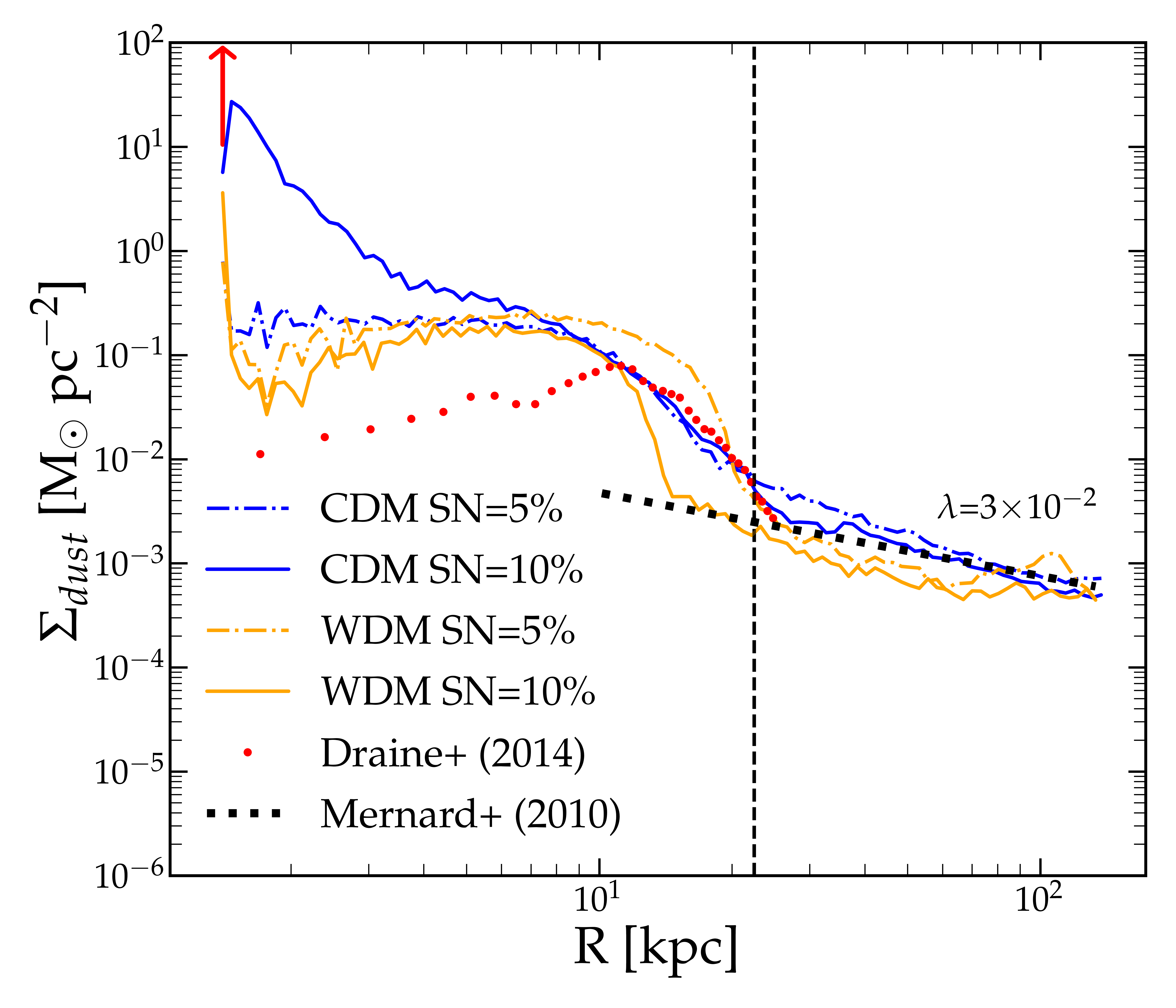}
    \caption{The dust surface density as a function of radius for all our simulations, at $z=0$. The grey dotted line is the scaling relation described in \protect\cite{menard_measuring_2010}, with an amplitude of 0.03, chosen to match the simulations at radii $\gtrsim$15kpc. The red points show the dust surface density from observations of M31 from \protect\cite{draine_andromedas_2014}. The vertical dashed line shows the masking radius of 15$h^{-1}$ckpc, and the red arrow shows the point at which gravitational forces become Newtonian.
    }
    \label{fig:dust_surf_dens}
\end{figure}

To better understand the difference in dust mass between the simulations, we plot the dust surface density as a function of the radius at $z=0$ in Fig. \ref{fig:dust_surf_dens}. We overplot the scaling relation $\Sigma_{\text{dust}}\propto r^{-0.8}$ from \cite{menard_measuring_2010} with a scaling amplitude $\lambda$=0.03 for the \dtm{} ratio, chosen to match the densities of the simulations at radii $\gtrsim 15$kpc, as well as a red arrow the point where gravity becomes Newtonian, and a dashed line at the masking radius. We also plot the radial dust surface density of M31, the closest Milky Way analogue to the Milky Way Galaxy. The surface density is taken from \cite{draine_andromedas_2014}, where they find a total estimated dust mass of $5.4\times10^{7}$M$_{\odot}$. Our results generally agree with the M31 observations in amplitude and shape at intermediate radii, between 10 and 20kpc, although inner regions have higher densities than M31. The high-density central regions are potentially inflated by limitations of our baryonic model; future refinements could add stellar winds or AGN feedback to help drive gas into the circumgalactic medium and lower the dust content in the central galaxy. The CDM simulation with 10 per cent supernova efficiency is additionally inflated by a recent close pass by a merging satellite, which happens at larger radii for the other CDM simulation and not at all in the WDM simulations. All simulations follow the \cite{menard_measuring_2010} scaling relation at radii greater than $\gtrsim$20kpc with little deviation. Outside of the central galaxy, all simulations have similar surface density profiles.

Appendix \ref{sec:app_dust_comp} shows the comparison between all dust mass prescriptions introduced in Section \ref{sec:dust_production}, where we find similar results and interpretations.

\subsection{Gini results}\label{sec:dust_evo_results}
\begin{figure}
    \includegraphics[width=\columnwidth]{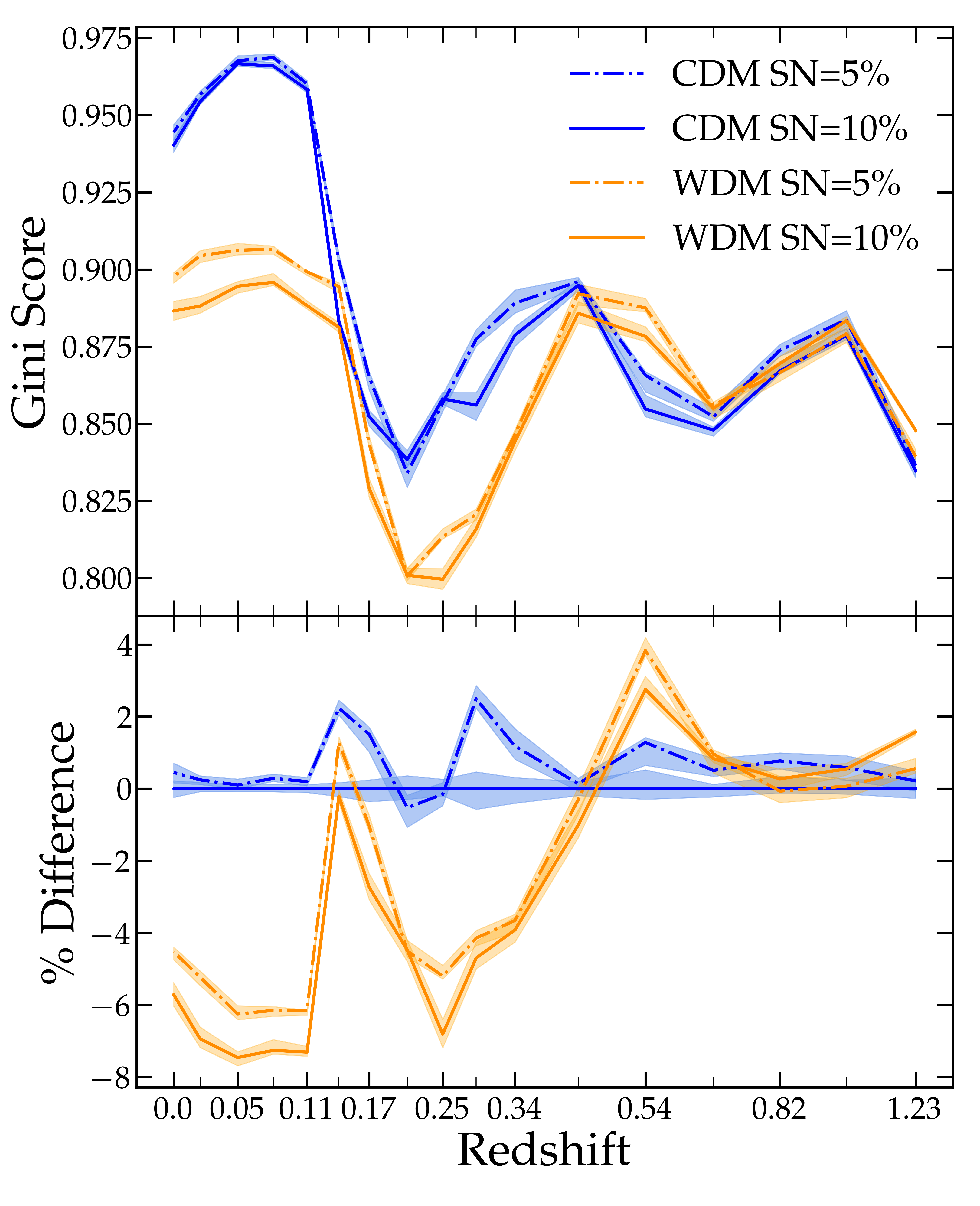}
    \caption{\textit{Top:} The Gini scores calculated from the dust distributions within the halo (e.g. Fig. \ref{fig:dust_hists_all}), where we mask the galaxy with a radius of 15$h^{-1}$ckpc, across all simulations from $z=1.22$ to $z=0$. The scores are calculated from a series of projections and then averaged for each snapshot. The shaded regions show 1$\sigma$ errors from a bootstrap resample of the 75 projections; the blue regions show errors for each CDM simulation, and the orange regions for each WDM simulation. \textit{Bottom:} The percentage differences between the Gini scores, normalised by the 10 per cent CDM simulation. Errors are the same as above.}
    \label{fig:gini_history}
\end{figure}

In Section \ref{sec:gini_intro}, we introduced the Gini score, which we now apply to our projected dust surface maps to further discriminate between their distributions and how this changes with time. The Gini score allows us to condense the information in maps like those in Figures \ref{fig:cdm_dust_hist} and \ref{fig:wdm_dust_hist}, into a single number. This is calculated for all halo rotations (see Section \ref{sec:gini_intro}) between $z=1.22$ and $z=0$, and is plotted in the top panel of Fig. \ref{fig:gini_history}. The shaded regions in the plot show 1$\sigma$ bootstrap errors for the medians at each redshift. The relative difference in Gini score with respect to the 10 per cent CDM simulation is shown in the bottom panel of Fig. \ref{fig:gini_history}, with the same errors as above. 

At higher redshift, cold and warm dark matter simulations share similar Gini scores and evolutions, but this begins to diverge around $z=0.5$. Both WDM simulations have lower concentrations of dust and hence a lower score than both CDM simulations, and the gap between them grows with time. Supernova feedback efficiency doesn’t appear to significantly impact the Gini score for a given dark matter type. The stochastic nature of the overall score is primarily governed by the merger history; for a universe such as WDM with fewer low mass haloes and fewer merger events, we expect reduced Gini scores compared to a purely CDM universe. This is a promising result as a proof of concept. Future work will focus on a wider range of Milky Way-like haloes and galaxy histories to explore the significance of the variance of such statistics.

\section{Discussion} \label{sec:Discussion}
There are several aspects of our results and analysis that require further discussion. In this section, we explore the main differences in our results between simulations, issues around measuring the Milky Way satellite population when drawing our conclusions, the impact of our choice of dust model in our hydrodynamic simulations, and finally, a discussion of the Gini statistic and how it is used in our analysis to draw conclusions. 

\subsection{Simulations} \label{sec:sims_discussion}
Our simulations and subsequent dust post-processing show that we have potentially uncovered a new observational probe to test the nature of dark matter. Through the use of varying supernova efficiencies as our lever, we created simulations from different underlying dark matter types that produce the same ``observable'' satellite populations. These different supernova efficiencies lead to different dust properties in each halo. We find that the total dust mass produced by the fiducial CDM simulation, with a 10 per cent supernova efficiency, is $\sim$4 times greater than in the fiducial WDM simulation. We also presented a preliminary exploration of how dust is spatially distributed within the halo using the Gini statistic, which shows significant and persistent differences between dark matter models. The difference between the two suggests that dust can be used as an observational probe of the underlying dark matter. In reality, galactic dust masses are influenced by many factors, such as star formation histories \citep{calzetti_dust_2000, clemens_dust_2013}, mergers \citep{buck_impact_2023}, and environments \citep{chelouche_dust_2007, vogelsberger_dust_2019, clark_quest_2023}, so more work must be done before we draw conclusions about any observationally favoured model. 

\subsection{Milky Way satellites}\label{sec:MW_sats}
There are many different sources for the mass of Milky Way satellites, which introduces some uncertainty into the correct values against which to calibrate the simulations. We use the masses published in \cite{lovell_matching_2023} for all satellites included in the comparison in Fig. \ref{fig:subhalo_central_pop}. This maintains a single source of data for simplicity and transparency, but does not explore the uncertainty of masses for these satellites. Many stellar masses are taken from \cite{mcconnachie_observed_2012}, which has minor differences with other published masses for these satellites \citep{woo_scaling_2008, fattahi_tidal_2018}, though all are quite consistent for the larger satellites. Additionally, there is some scatter in the dynamical masses compared with \cite{strigari_common_2008} and \cite{wolf_accurate_2010}, but once again, being the largest Milky Way satellites, the scatter in these measurements is relatively small.

Counter-intuitively, Fig. \ref{fig:subhalo_central_pop} shows that the fiducial WDM simulation produces more low-mass satellite galaxies than any other simulation, as mentioned in Section \ref{sec:tuning_results}. This is rather unexpected, given that WDM is generally implemented to decrease satellite numbers. The apparent overproduction in the simulation \footnote{Although we note that both WDM simulations produce fewer \textit{total subhaloes} than their CDM counterparts.} could be due to a number of factors, first among them the artificial fragmentation common in WDM simulations \citep{wang_discreteness_2007}. We have explored removing such artificial haloes by implementing a minimum 100-particle limit. Our testing suggests that artificial haloes are not the cause of the excess. More work is needed to identify spurious haloes, e.g. identifying pancake haloes in the initial conditions \citep{lovell_properties_2014}, but this is beyond the scope of this paper. Alternatively, \cite{oman_warm_2024} found that WDM may boost the number of low-mass galaxies in 21-cm hydrogen observations. They state this may be because galaxies tend to form later in WDM simulations, and the subsequent delay in star formation helps them retain more gas than their CDM counterparts. Additionally, the higher concentration of CDM haloes may drive faster merging times, causing an excess of WDM satellites that have yet to fully merge by the same point in time. This may be true, at least for our Milky Way analogue. We defer such an investigation to later work, conscious that the most stringent limits on WDM particle masses are based on the observation of satellites below the resolution of these simulations \citep{banik_novel_2021, nadler_constraints_2021, dekker_warm_2022}. However, the possibility of a WDM universe potentially producing more satellite galaxies in some haloes prompts further investigation into the variety of pathways for satellite accretion in a WDM universe and how these might compare with our own Milky Way satellites.

\subsection{Dust} \label{sec:dust_disscusion}
As the goal of this paper is to provide a simple proof of concept, we chose the simplest \powderday{} model to generate dust from our hydrodynamic simulations, the \dtm{} ratio. This correlation has a long track record in the literature \citep{silva_modeling_1998, james_scuba_2002, lacey_galaxy_2008, somerville_galaxy_2012, yajima_observational_2015, ma_dust_2019}. While there are more detailed implementations of dust scaling relations and more advanced subgrid models that treat dust evolution \citep[e.g. ][]{mckinnon_dust_2016, aoyama_cosmological_2018}, we make this choice to keep the comparisons between simulations and the observations straightforward and transparent.  However, we tested two other dust mass prescriptions included in Powderday to confirm the results from the \dtm{} ratio. 

These two models were the \cite{remy-ruyer_gas--dust_2014} gas-to-dust ratio and the \cite{li_dust--gas_2019} best-fitting set of parameters. In short, the \dtm{} ratio produces the least total dust in all simulations, and the best-fitting model produces the most dust in all simulations. All prescriptions produce highly concentrated dust distributions, though the transition in the gas-to-dust and best-fitting prescriptions is more extreme from high to low densities. Despite these differences in the spatial distribution between dust mass prescriptions, we have confirmed they do not significantly change the Gini scores. Further comparison of the dust models is provided in Appendix \ref{sec:app_dust_comp}.

\subsection{Gini trends}\label{sec:gini_discussion}
The difference in the Gini evolution produced by our simulations is an encouraging sign that the dust around a galaxy contains spatial information that can be used to differentiate between dark matter models. The individual Gini scores appear to be governed more strongly by the halo evolution rather than the baryonic model, which has two primary implications. First, the diffuse dust is more dependent on the amount of substructure than the baryonic model. This is impacted by masking the central galaxy, where most of the dust is concentrated. For example, the gas disc is much more extended in the 5 per cent WDM simulation in Fig. \ref{fig:wdm_dust_hist} compared to the 10 per cent CDM simulation in Fig. \ref{fig:cdm_dust_hist}. By masking this region to avoid saturating the Gini signal, we are also masking the region with the greatest difference caused by our implemented baryonic model. 

The second implication is that every halo in our simulation has a unique evolution, even when starting from the same initial conditions and random seed. This unique evolution makes direct extrapolation of a Gini score to the observations difficult. However, the relative difference in the simulations highlights a difference that can be leveraged in observations. The absolute values for the total dust mass and Gini scores will likely change as a function of the random initial seed. Still, we expect that a WDM universe, with fewer average mergers, will produce lower Gini scores over a large sample. Any prediction for an average Gini score in dust observations would require many simulations and careful treatment of the comparison between the two; though not beyond the capabilities of modern telescopes, with JWST and ALMA observations able to achieve sub-kpc resolution at $z\approx1$ \citep[e.g.][]{liu_jwst_2024}. This is beyond the scope of the current work but will be the focus of our future work.

We see modest changes in the absolute Gini score across all prescriptions, but little change in the shape of the evolution. This is explored further in Appendix \ref{sec:app_dust_comp}.

\section{Conclusions} \label{sec:Conclusions}
We have tested a novel method for distinguishing between cold and warm dark matter using the dust around a galaxy. We ran zoom-in simulations of Milky Way analogues, tuning the efficiency of the supernova feedback to create systems with degenerate satellite galaxy populations. The different supernova efficiencies create different dust populations, which are used as the observational probe between dark matter models.

Our simulations take identical random seeds to create the initial conditions. The parent dark matter only simulations are then used to identify Milky Way analogues, hosted by haloes with a mass of approximately $2\times10^{12}$\msun{}. Once identified, we ensure the haloes are isolated, to make sure there is not another large halo within 8 Mpc. After identifying target haloes in both the cold and warm dark matter parent simulations, the zoom-in simulations were run with 5 per cent and 10 per cent supernova efficiencies. The resimulated Milky Way analogue galaxies all show general agreement with observed quantities of the real Milky Way (see Table \ref{tab:sim_mass_summary}).

We post-processed each simulation using \powderday{} \citep{narayanan_powderday_2021} to produce the dust properties, based on the simulated gas particle properties. The results presented here use a \dtm{} ratio of 0.4 \citep{dwek_evolution_1998,yajima_observational_2015}, and we confirmed they are not overly sensitive to the precise dust prescription. \powderday{} produces distinct dust properties for each simulation, from which the main results from the dust analysis are as follows.

\begin{enumerate}
    \item The different supernova efficiencies produce different dust masses; the fiducial CDM simulation produces the most at $4.56\times10^{7}$ M$_{\odot}$, compared to $1.05\times10^{7}$ M$_{\odot}$ in the fiducial WDM simulation. All dust masses are listed in Table \ref{tab:dust_masses}.
    
    \item The majority of the dust is always contained within the inner 15kpc of the halo, evident in Fig. \ref{fig:dust_hists_all}. We see a generally higher dust mass with CDM across much of the halo (Fig. \ref{fig:dust_diff}). There are regions with obvious differences, which are largely generated by differences in satellite positioning and an extended galactic disc in the WDM simulation.
    
    \item The dust surface density profiles seen in Fig. \ref{fig:dust_surf_dens} show good agreement with the  \cite{menard_measuring_2010} scaling relation, $\Sigma_{\text{dust}}\propto r^{-0.8}$, at radii larger than $\sim$20kpc. Between 10 and 20kpc, most simulations match reasonably well with the \cite{draine_andromedas_2014} dust observations of M31, though, below these scales, the fiducial CDM simulation has a rising density profile, whereas both WDM simulations and the 5 per cent CDM simulation have flat density profiles. 
    
    \item We calculated the Gini coefficient of each dust distribution, masking the inner 15kpc to study the diffuse dust, in Fig. \ref{fig:gini_history}. We compared the evolution of the dust concentration across all four simulations, finding that a WDM universe produces lower concentrations of dust. The evolution of the dust concentration is impacted more so by the merger history than the supernova efficiency, indicating these sorts of spatial studies can be leveraged by observations of dust around Milky Way-type galaxies.
     
\end{enumerate}

Our results show that cold and warm dark matter simulations can produce similar satellite galaxy populations with different dust distributions. The total dust mass is different across all simulations, and there is a clear separation in the evolution of the masked Gini scores for these galaxies. There are two clear avenues for future work. The first is to use more advanced baryonic models, potentially with self-consistent dust treatment. As previously mentioned in Section \ref{sec:baryonic_model}, the baryonic models used are simplistic compared to current flagship simulations. However, using more complex models requires more effort to disentangle the effects of the subgrid routines from the different dark matter models. A more straightforward approach is to first increase the sample of galaxies used in the Gini analysis in an effort to overcome the expected variation from different random seeds. We can then compare the results with observational catalogues such as DustPedia \citep{davies_dustpedia_2017} using simulated images generated by \powderday{}. The results from this and future work pave the way for developing a new tool for determining the nature of dark matter in our Universe. 

\section*{Acknowledgements}
The authors would like to thank the referee, Mark Lo v ell, for their valuable comments in improving this publication, as well as useful discussions with Alexander Knebe and Desika Narayanan. AU and RM also acknowledge the University of Western Australia/ICRAR for supporting a research visit. This research was supported in part by the Australian Government through the Australian Research Council Centre of Excellence for Dark Matter Particle Physics (CDM, CE200100008), and in part by the Australian Research Council Centre of Excellence for All Sky Astrophysics in 3 Dimensions (ASTRO 3D, CE170100013). AU acknowledges the support of an Australian Government Research Training Program Scholarship, and DC acknowledges the support of an ARC Future Fellowship (FT220100841). This work was performed on the OzSTAR national facility at Swinburne University of Technology. The OzSTAR program receives funding in part from the Astronomy National Collaborative Research Infrastructure Strategy (NCRIS) allocation provided by the Australian Government, and from the Victorian Higher Education State Investment Fund (VHESIF) provided by the Victorian Government.

%%%%%%%%%%%%%%%%%%%%%%%%%%%%%%%%%%%%%%%%%%%%%%%%%%
\section*{Data Availability}
The simulations used in this study are available and accessible on reasonable request. The data products of this work can be shared upon request to the corresponding author.



\bibliographystyle{mnras}
\bibliography{references}

\begin{thebibliography}{}
\makeatletter
\relax
\def\mn@urlcharsother{\let\do\@makeother \do\$\do\&\do\#\do\^\do\_\do\%\do\~}
\def\mn@doi{\begingroup\mn@urlcharsother \@ifnextchar [ {\mn@doi@} {\mn@doi@[]}}
\def\mn@doi@[#1]#2{\def\@tempa{#1}\ifx\@tempa\@empty \href {http://dx.doi.org/#2} {doi:#2}\else \href {http://dx.doi.org/#2} {#1}\fi \endgroup}
\def\mn@eprint#1#2{\mn@eprint@#1:#2::\@nil}
\def\mn@eprint@arXiv#1{\href {http://arxiv.org/abs/#1} {{\tt arXiv:#1}}}
\def\mn@eprint@dblp#1{\href {http://dblp.uni-trier.de/rec/bibtex/#1.xml} {dblp:#1}}
\def\mn@eprint@#1:#2:#3:#4\@nil{\def\@tempa {#1}\def\@tempb {#2}\def\@tempc {#3}\ifx \@tempc \@empty \let \@tempc \@tempb \let \@tempb \@tempa \fi \ifx \@tempb \@empty \def\@tempb {arXiv}\fi \@ifundefined {mn@eprint@\@tempb}{\@tempb:\@tempc}{\expandafter \expandafter \csname mn@eprint@\@tempb\endcsname \expandafter{\@tempc}}}

\bibitem[\protect\citeauthoryear{Aalbers et~al.,}{Aalbers et~al.}{2023}]{aalbers_first_2023}
Aalbers J.,  et~al., 2023, \mn@doi [Physical Review Letters] {10.1103/PhysRevLett.131.041002}, 131, 041002

\bibitem[\protect\citeauthoryear{Abraham, van~den Bergh  \& Nair}{Abraham et~al.}{2003}]{abraham_new_2003}
Abraham R.~G.,  van~den Bergh S.,   Nair P.,  2003, \mn@doi [The Astrophysical Journal] {10.1086/373919}, 588, 218

\bibitem[\protect\citeauthoryear{Adhikari et~al.,}{Adhikari et~al.}{2021}]{adhikari_strong_2021}
Adhikari G.,  et~al., 2021, \mn@doi [Science Advances] {10.1126/sciadv.abk2699}, 7, eabk2699

\bibitem[\protect\citeauthoryear{Aghanim et~al.,}{Aghanim et~al.}{2020}]{aghanim_planck_2020}
Aghanim N.,  et~al., 2020, \mn@doi [Astronomy and Astrophysics] {10.1051/0004-6361/201833910}, 641, A6

\bibitem[\protect\citeauthoryear{Aguirre}{Aguirre}{1999}]{aguirre_intergalactic_1999}
Aguirre A.,  1999, \mn@doi [The Astrophysical Journal] {10.1086/307945}, 525, 583

\bibitem[\protect\citeauthoryear{Aoyama, Hou, Hirashita, Nagamine  \& Shimizu}{Aoyama et~al.}{2018}]{aoyama_cosmological_2018}
Aoyama S.,  Hou K.-C.,  Hirashita H.,  Nagamine K.,   Shimizu I.,  2018, \mn@doi [Monthly Notices of the Royal Astronomical Society] {10.1093/mnras/sty1431}, 478, 4905

\bibitem[\protect\citeauthoryear{Aprile et~al.,}{Aprile et~al.}{2023}]{aprile_first_2023}
Aprile E.,  et~al., 2023, \mn@doi [Physical Review Letters] {10.1103/PhysRevLett.131.041003}, 131, 041003

\bibitem[\protect\citeauthoryear{Avila-Reese, Colín, Valenzuela, D'Onghia  \& Firmani}{Avila-Reese et~al.}{2001}]{avila-reese_formation_2001}
Avila-Reese V.,  Colín P.,  Valenzuela O.,  D'Onghia E.,   Firmani C.,  2001, \mn@doi [The Astrophysical Journal] {10.1086/322411}, 559, 516

\bibitem[\protect\citeauthoryear{Bagla}{Bagla}{2002}]{bagla_treepm_2002}
Bagla J.~S.,  2002, \mn@doi [Journal of Astrophysics and Astronomy] {10.1007/BF02702282}, 23, 185

\bibitem[\protect\citeauthoryear{Banik, Bovy, Bertone, Erkal  \& de Boer}{Banik et~al.}{2021}]{banik_novel_2021}
Banik N.,  Bovy J.,  Bertone G.,  Erkal D.,   de Boer T. J.~L.,  2021, \mn@doi [Journal of Cosmology and Astroparticle Physics] {10.1088/1475-7516/2021/10/043}, 2021, 043

\bibitem[\protect\citeauthoryear{Benito, Criado, Hütsi, Raidal  \& Veermäe}{Benito et~al.}{2020}]{benito_implications_2020}
Benito M.,  Criado J.~C.,  Hütsi G.,  Raidal M.,   Veermäe H.,  2020, \mn@doi [Physical Review D] {10.1103/PhysRevD.101.103023}, 101, 103023

\bibitem[\protect\citeauthoryear{Benson, Frenk, Lacey, Baugh  \& Cole}{Benson et~al.}{2002}]{benson_effects_2002}
Benson A.~J.,  Frenk C.~S.,  Lacey C.~G.,  Baugh C.~M.,   Cole S.,  2002, \mn@doi [Monthly Notices of the Royal Astronomical Society] {10.1046/j.1365-8711.2002.05388.x}, 333, 177

\bibitem[\protect\citeauthoryear{Bernabei et~al.,}{Bernabei et~al.}{2000}]{bernabei_search_2000}
Bernabei R.,  et~al., 2000, \mn@doi [Physics Letters B] {10.1016/S0370-2693(00)00405-6}, 480, 23

\bibitem[\protect\citeauthoryear{Bernabei et~al.,}{Bernabei et~al.}{2008}]{bernabei_first_2008}
Bernabei R.,  et~al., 2008, \mn@doi [European Physical Journal C] {10.1140/epjc/s10052-008-0662-y}, 56, 333

\bibitem[\protect\citeauthoryear{Bernabei et~al.,}{Bernabei et~al.}{2021}]{bernabei_dark_2021}
Bernabei R.,  et~al., 2021, The dark matter: {DAMA}/{LIBRA} and its perspectives, \url {http://arxiv.org/abs/2110.04734}

\bibitem[\protect\citeauthoryear{Bianchi \& Schneider}{Bianchi \& Schneider}{2007}]{bianchi_dust_2007}
Bianchi S.,  Schneider R.,  2007, \mn@doi [Monthly Notices of the Royal Astronomical Society] {10.1111/j.1365-2966.2007.11829.x}, 378, 973

\bibitem[\protect\citeauthoryear{Bland-Hawthorn \& Gerhard}{Bland-Hawthorn \& Gerhard}{2016}]{bland-hawthorn_galaxy_2016}
Bland-Hawthorn J.,  Gerhard O.,  2016, \mn@doi [Annual Review of Astronomy and Astrophysics] {10.1146/annurev-astro-081915-023441}, 54, 529

\bibitem[\protect\citeauthoryear{Bode, Ostriker  \& Turok}{Bode et~al.}{2001}]{bode_halo_2001}
Bode P.,  Ostriker J.~P.,   Turok N.,  2001, \mn@doi [The Astrophysical Journal] {10.1086/321541}, 556, 93

\bibitem[\protect\citeauthoryear{Boehm \& Schaeffer}{Boehm \& Schaeffer}{2005}]{boehm_constraints_2005}
Boehm C.,  Schaeffer R.,  2005, \mn@doi [Astronomy and Astrophysics] {10.1051/0004-6361:20042238}, 438, 419

\bibitem[\protect\citeauthoryear{Boehm, Schewtschenko, Wilkinson, Baugh  \& Pascoli}{Boehm et~al.}{2014}]{boehm_using_2014}
Boehm C.,  Schewtschenko J.~A.,  Wilkinson R.~J.,  Baugh C.~M.,   Pascoli S.,  2014, \mn@doi [Monthly Notices of the Royal Astronomical Society] {10.1093/mnrasl/slu115}, 445, L31

\bibitem[\protect\citeauthoryear{Boylan-Kolchin, Bullock  \& Kaplinghat}{Boylan-Kolchin et~al.}{2012}]{boylan-kolchin_milky_2012}
Boylan-Kolchin M.,  Bullock J.~S.,   Kaplinghat M.,  2012, \mn@doi [Monthly Notices of the Royal Astronomical Society] {10.1111/j.1365-2966.2012.20695.x}, 422, 1203

\bibitem[\protect\citeauthoryear{Buck, Obreja, Ratcliffe, Lu, Minchev  \& Macciò}{Buck et~al.}{2023}]{buck_impact_2023}
Buck T.,  Obreja A.,  Ratcliffe B.,  Lu Y.,  Minchev I.,   Macciò A.~V.,  2023, \mn@doi [Monthly Notices of the Royal Astronomical Society] {10.1093/mnras/stad1503}, 523, 1565

\bibitem[\protect\citeauthoryear{Bullock, Kravtsov  \& Weinberg}{Bullock et~al.}{2000}]{bullock_reionization_2000}
Bullock J.~S.,  Kravtsov A.~V.,   Weinberg D.~H.,  2000, \mn@doi [The Astrophysical Journal] {10.1086/309279}, 539, 517

\bibitem[\protect\citeauthoryear{Calzetti, Armus, Bohlin, Kinney, Koornneef  \& Storchi-Bergmann}{Calzetti et~al.}{2000}]{calzetti_dust_2000}
Calzetti D.,  Armus L.,  Bohlin R.~C.,  Kinney A.~L.,  Koornneef J.,   Storchi-Bergmann T.,  2000, \mn@doi [The Astrophysical Journal] {10.1086/308692}, 533, 682

\bibitem[\protect\citeauthoryear{Chaikin, Schaye, Schaller, Benítez-Llambay, Nobels  \& Ploeckinger}{Chaikin et~al.}{2023}]{chaikin_thermal-kinetic_2023}
Chaikin E.,  Schaye J.,  Schaller M.,  Benítez-Llambay A.,  Nobels F. S.~J.,   Ploeckinger S.,  2023, \mn@doi [Monthly Notices of the Royal Astronomical Society] {10.1093/mnras/stad1626}, 523, 3709

\bibitem[\protect\citeauthoryear{Chelouche, Koester  \& Bowen}{Chelouche et~al.}{2007}]{chelouche_dust_2007}
Chelouche D.,  Koester B.~P.,   Bowen D.~V.,  2007, \mn@doi [The Astrophysical Journal] {10.1086/525251}, 671, L97

\bibitem[\protect\citeauthoryear{Chomiuk \& Povich}{Chomiuk \& Povich}{2011}]{chomiuk_toward_2011}
Chomiuk L.,  Povich M.~S.,  2011, \mn@doi [The Astronomical Journal] {10.1088/0004-6256/142/6/197}, 142, 197

\bibitem[\protect\citeauthoryear{Clark, Roman-Duval, Gordon, Bot, Smith  \& Hagen}{Clark et~al.}{2023}]{clark_quest_2023}
Clark C. J.~R.,  Roman-Duval J.~C.,  Gordon K.~D.,  Bot C.,  Smith M. W.~L.,   Hagen L. M.~Z.,  2023, \mn@doi [The Astrophysical Journal] {10.3847/1538-4357/acbb66}, 946, 42

\bibitem[\protect\citeauthoryear{Clemens et~al.,}{Clemens et~al.}{2013}]{clemens_dust_2013}
Clemens M.~S.,  et~al., 2013, \mn@doi [Monthly Notices of the Royal Astronomical Society] {10.1093/mnras/stt760}, 433, 695

\bibitem[\protect\citeauthoryear{Clowe, Bradač, Gonzalez, Markevitch, Randall, Jones  \& Zaritsky}{Clowe et~al.}{2006}]{clowe_direct_2006}
Clowe D.,  Bradač M.,  Gonzalez A.~H.,  Markevitch M.,  Randall S.~W.,  Jones C.,   Zaritsky D.,  2006, \mn@doi [The Astrophysical Journal] {10.1086/508162}, 648, L109

\bibitem[\protect\citeauthoryear{Coarasa et~al.,}{Coarasa et~al.}{2024}]{coarasa_anais-112_2024}
Coarasa I.,  et~al., 2024, {ANAIS}-112 three years data: a sensitive model independent negative test of the {DAMA}/{LIBRA} dark matter signal, \mn@doi{10.48550/arXiv.2404.17348}, \url {https://ui.adsabs.harvard.edu/abs/2024arXiv240417348C}

\bibitem[\protect\citeauthoryear{Colín, Valenzuela  \& Avila-Reese}{Colín et~al.}{2008}]{colin_structure_2008}
Colín P.,  Valenzuela O.,   Avila-Reese V.,  2008, \mn@doi [The Astrophysical Journal] {10.1086/524030}, 673, 203

\bibitem[\protect\citeauthoryear{Conroy \& Gunn}{Conroy \& Gunn}{2010}]{conroy_propagation_2010}
Conroy C.,  Gunn J.~E.,  2010, \mn@doi [The Astrophysical Journal] {10.1088/0004-637X/712/2/833}, 712, 833

\bibitem[\protect\citeauthoryear{Crain et~al.,}{Crain et~al.}{2015}]{crain_eagle_2015}
Crain R.~A.,  et~al., 2015, \mn@doi [Monthly Notices of the Royal Astronomical Society] {10.1093/mnras/stv725}, 450, 1937

\bibitem[\protect\citeauthoryear{Croton et~al.,}{Croton et~al.}{2006}]{croton_many_2006}
Croton D.~J.,  et~al., 2006, \mn@doi [Monthly Notices of the Royal Astronomical Society] {10.1111/j.1365-2966.2005.09675.x}, 365, 11

\bibitem[\protect\citeauthoryear{Dalla~Vecchia \& Schaye}{Dalla~Vecchia \& Schaye}{2008}]{dalla_vecchia_simulating_2008}
Dalla~Vecchia C.,  Schaye J.,  2008, \mn@doi [Monthly Notices of the Royal Astronomical Society] {10.1111/j.1365-2966.2008.13322.x}, 387, 1431

\bibitem[\protect\citeauthoryear{Davies et~al.,}{Davies et~al.}{2017}]{davies_dustpedia_2017}
Davies J.~I.,  et~al., 2017, \mn@doi [Publications of the Astronomical Society of the Pacific] {10.1088/1538-3873/129/974/044102}, 129, 044102

\bibitem[\protect\citeauthoryear{Davis, Efstathiou, Frenk  \& White}{Davis et~al.}{1985}]{davis_evolution_1985}
Davis M.,  Efstathiou G.,  Frenk C.~S.,   White S. D.~M.,  1985, \mn@doi [The Astrophysical Journal] {10.1086/163168}, 292, 371

\bibitem[\protect\citeauthoryear{Davé, Hernquist, Katz  \& Weinberg}{Davé et~al.}{1999}]{dave_low-redshift_1999}
Davé R.,  Hernquist L.,  Katz N.,   Weinberg D.~H.,  1999, \mn@doi [The Astrophysical Journal] {10.1086/306722}, 511, 521

\bibitem[\protect\citeauthoryear{Davé, Anglés-Alcázar, Narayanan, Li, Rafieferantsoa  \& Appleby}{Davé et~al.}{2019}]{dave_simba_2019}
Davé R.,  Anglés-Alcázar D.,  Narayanan D.,  Li Q.,  Rafieferantsoa M.~H.,   Appleby S.,  2019, \mn@doi [Monthly Notices of the Royal Astronomical Society] {10.1093/mnras/stz937}, 486, 2827

\bibitem[\protect\citeauthoryear{Dekker, Ando, Correa  \& Ng}{Dekker et~al.}{2022}]{dekker_warm_2022}
Dekker A.,  Ando S.,  Correa C.~A.,   Ng K. C.~Y.,  2022, \mn@doi [Physical Review D] {10.1103/PhysRevD.106.123026}, 106, 123026

\bibitem[\protect\citeauthoryear{Draine et~al.,}{Draine et~al.}{2014}]{draine_andromedas_2014}
Draine B.~T.,  et~al., 2014, \mn@doi [The Astrophysical Journal] {10.1088/0004-637X/780/2/172}, 780, 172

\bibitem[\protect\citeauthoryear{Dwek}{Dwek}{1998}]{dwek_evolution_1998}
Dwek E.,  1998, \mn@doi [The Astrophysical Journal] {10.1086/305829}, 501, 643

\bibitem[\protect\citeauthoryear{Eisenstein \& Hu}{Eisenstein \& Hu}{1998}]{eisenstein_baryonic_1998}
Eisenstein D.~J.,  Hu W.,  1998, \mn@doi [The Astrophysical Journal] {10.1086/305424}, 496, 605

\bibitem[\protect\citeauthoryear{Eisenstein et~al.,}{Eisenstein et~al.}{2005}]{eisenstein_detection_2005}
Eisenstein D.~J.,  et~al., 2005, \mn@doi [The Astrophysical Journal] {10.1086/466512}, 633, 560

\bibitem[\protect\citeauthoryear{Elia et~al.,}{Elia et~al.}{2022}]{elia_star_2022}
Elia D.,  et~al., 2022, \mn@doi [The Astrophysical Journal] {10.3847/1538-4357/aca27d}, 941, 162

\bibitem[\protect\citeauthoryear{Fattahi, Navarro, Frenk, Oman, Sawala  \& Schaller}{Fattahi et~al.}{2018}]{fattahi_tidal_2018}
Fattahi A.,  Navarro J.~F.,  Frenk C.~S.,  Oman K.~A.,  Sawala T.,   Schaller M.,  2018, \mn@doi [Monthly Notices of the Royal Astronomical Society] {10.1093/mnras/sty408}, 476, 3816

\bibitem[\protect\citeauthoryear{Goodenough \& Hooper}{Goodenough \& Hooper}{2009}]{goodenough_possible_2009}
Goodenough L.,  Hooper D.,  2009, Possible {Evidence} {For} {Dark} {Matter} {Annihilation} {In} {The} {Inner} {Milky} {Way} {From} {The} {Fermi} {Gamma} {Ray} {Space} {Telescope}, \mn@doi{10.48550/arXiv.0910.2998}, \url {https://ui.adsabs.harvard.edu/abs/2009arXiv0910.2998G}

\bibitem[\protect\citeauthoryear{Governato, Willman, Mayer, Brooks, Stinson, Valenzuela, Wadsley  \& Quinn}{Governato et~al.}{2007}]{governato_forming_2007}
Governato F.,  Willman B.,  Mayer L.,  Brooks A.,  Stinson G.,  Valenzuela O.,  Wadsley J.,   Quinn T.,  2007, \mn@doi [Monthly Notices of the Royal Astronomical Society] {10.1111/j.1365-2966.2006.11266.x}, 374, 1479

\bibitem[\protect\citeauthoryear{Haardt \& Madau}{Haardt \& Madau}{1996}]{haardt_radiative_1996}
Haardt F.,  Madau P.,  1996, \mn@doi [The Astrophysical Journal] {10.1086/177035}, 461, 20

\bibitem[\protect\citeauthoryear{Hahn \& Abel}{Hahn \& Abel}{2011}]{hahn_multi-scale_2011}
Hahn O.,  Abel T.,  2011, \mn@doi [Monthly Notices of the Royal Astronomical Society] {10.1111/j.1365-2966.2011.18820.x}, 415, 2101

\bibitem[\protect\citeauthoryear{Han, Cole, Frenk, Benitez-Llambay  \& Helly}{Han et~al.}{2018}]{han_hbt_2018}
Han J.,  Cole S.,  Frenk C.~S.,  Benitez-Llambay A.,   Helly J.,  2018, \mn@doi [Monthly Notices of the Royal Astronomical Society] {10.1093/mnras/stx2792}, 474, 604

\bibitem[\protect\citeauthoryear{Hernández-Aguayo et~al.,}{Hernández-Aguayo et~al.}{2022}]{hernandez-aguayo_millenniumtng_2022}
Hernández-Aguayo C.,  et~al., 2022, The {MillenniumTNG} {Project}: {High}-precision predictions for matter clustering and halo statistics, \mn@doi{10.48550/arXiv.2210.10059}, \url {http://arxiv.org/abs/2210.10059}

\bibitem[\protect\citeauthoryear{Hooper \& Goodenough}{Hooper \& Goodenough}{2011}]{hooper_dark_2011}
Hooper D.,  Goodenough L.,  2011, \mn@doi [Physics Letters B] {10.1016/j.physletb.2011.02.029}, 697, 412

\bibitem[\protect\citeauthoryear{James, Dunne, Eales  \& Edmunds}{James et~al.}{2002}]{james_scuba_2002}
James A.,  Dunne L.,  Eales S.,   Edmunds M.~G.,  2002, \mn@doi [Monthly Notices of the Royal Astronomical Society] {10.1046/j.1365-8711.2002.05660.x}, 335, 753

\bibitem[\protect\citeauthoryear{Kafle, Sharma, Lewis  \& Bland-Hawthorn}{Kafle et~al.}{2014}]{kafle_shoulders_2014}
Kafle P.~R.,  Sharma S.,  Lewis G.~F.,   Bland-Hawthorn J.,  2014, \mn@doi [The Astrophysical Journal] {10.1088/0004-637X/794/1/59}, 794, 59

\bibitem[\protect\citeauthoryear{Katz, Weinberg  \& Hernquist}{Katz et~al.}{1996}]{katz_cosmological_1996}
Katz N.,  Weinberg D.~H.,   Hernquist L.,  1996, \mn@doi [The Astrophysical Journal Supplement Series] {10.1086/192305}, 105, 19

\bibitem[\protect\citeauthoryear{Kaviraj et~al.,}{Kaviraj et~al.}{2017}]{kaviraj_horizon-agn_2017}
Kaviraj S.,  et~al., 2017, \mn@doi [Monthly Notices of the Royal Astronomical Society] {10.1093/mnras/stx126}, 467, 4739

\bibitem[\protect\citeauthoryear{Klasen, Pohl  \& Sigl}{Klasen et~al.}{2015}]{klasen_indirect_2015}
Klasen M.,  Pohl M.,   Sigl G.,  2015, \mn@doi [Progress in Particle and Nuclear Physics] {10.1016/j.ppnp.2015.07.001}, 85, 1

\bibitem[\protect\citeauthoryear{Klypin, Holtzman, Primack  \& Regos}{Klypin et~al.}{1993}]{klypin_structure_1993}
Klypin A.,  Holtzman J.,  Primack J.,   Regos E.,  1993, \mn@doi [The Astrophysical Journal] {10.1086/173210}, 416, 1

\bibitem[\protect\citeauthoryear{Klypin, Kravtsov, Valenzuela  \& Prada}{Klypin et~al.}{1999}]{klypin_where_1999}
Klypin A.~A.,  Kravtsov A.~V.,  Valenzuela O.,   Prada F.,  1999, \mn@doi [The Astrophysical Journal] {10.1086/307643}, 522, 82

\bibitem[\protect\citeauthoryear{Lacey, Baugh, Frenk, Silva, Granato  \& Bressan}{Lacey et~al.}{2008}]{lacey_galaxy_2008}
Lacey C.~G.,  Baugh C.~M.,  Frenk C.~S.,  Silva L.,  Granato G.~L.,   Bressan A.,  2008, \mn@doi [Monthly Notices of the Royal Astronomical Society] {10.1111/j.1365-2966.2008.12949.x}, 385, 1155

\bibitem[\protect\citeauthoryear{Li, Narayanan  \& Davé}{Li et~al.}{2019}]{li_dust--gas_2019}
Li Q.,  Narayanan D.,   Davé R.,  2019, \mn@doi [Monthly Notices of the Royal Astronomical Society] {10.1093/mnras/stz2684}, 490, 1425

\bibitem[\protect\citeauthoryear{Licquia \& Newman}{Licquia \& Newman}{2015}]{licquia_improved_2015}
Licquia T.~C.,  Newman J.~A.,  2015, \mn@doi [The Astrophysical Journal] {10.1088/0004-637X/806/1/96}, 806, 96

\bibitem[\protect\citeauthoryear{Liu et~al.,}{Liu et~al.}{2024}]{liu_jwst_2024}
Liu Z.,  et~al., 2024, \mn@doi [The Astrophysical Journal] {10.3847/1538-4357/ad4096}, 968, 15

\bibitem[\protect\citeauthoryear{Lotz, Primack  \& Madau}{Lotz et~al.}{2004}]{lotz_new_2004}
Lotz J.~M.,  Primack J.,   Madau P.,  2004, \mn@doi [The Astronomical Journal] {10.1086/421849}, 128, 163

\bibitem[\protect\citeauthoryear{Lovell \& Zavala}{Lovell \& Zavala}{2023}]{lovell_matching_2023}
Lovell M.~R.,  Zavala J.,  2023, \mn@doi [Monthly Notices of the Royal Astronomical Society] {10.1093/mnras/stad216}, 520, 1567

\bibitem[\protect\citeauthoryear{Lovell, Frenk, Eke, Jenkins, Gao  \& Theuns}{Lovell et~al.}{2014}]{lovell_properties_2014}
Lovell M.~R.,  Frenk C.~S.,  Eke V.~R.,  Jenkins A.,  Gao L.,   Theuns T.,  2014, \mn@doi [Monthly Notices of the Royal Astronomical Society] {10.1093/mnras/stt2431}, 439, 300

\bibitem[\protect\citeauthoryear{Ma et~al.,}{Ma et~al.}{2019}]{ma_dust_2019}
Ma X.,  et~al., 2019, \mn@doi [Monthly Notices of the Royal Astronomical Society] {10.1093/mnras/stz1324}, 487, 1844

\bibitem[\protect\citeauthoryear{McConnachie}{McConnachie}{2012}]{mcconnachie_observed_2012}
McConnachie A.~W.,  2012, \mn@doi [The Astronomical Journal] {10.1088/0004-6256/144/1/4}, 144, 4

\bibitem[\protect\citeauthoryear{McConnachie \& Venn}{McConnachie \& Venn}{2020}]{mcconnachie_updated_2020}
McConnachie A.~W.,  Venn K.~A.,  2020, \mn@doi [Research Notes of the American Astronomical Society] {10.3847/2515-5172/abd18b}, 4, 229

\bibitem[\protect\citeauthoryear{McKinnon, Torrey  \& Vogelsberger}{McKinnon et~al.}{2016}]{mckinnon_dust_2016}
McKinnon R.,  Torrey P.,   Vogelsberger M.,  2016, \mn@doi [Monthly Notices of the Royal Astronomical Society] {10.1093/mnras/stw253}, 457, 3775

\bibitem[\protect\citeauthoryear{McMillan}{McMillan}{2017}]{mcmillan_mass_2017}
McMillan P.~J.,  2017, \mn@doi [Monthly Notices of the Royal Astronomical Society] {10.1093/mnras/stw2759}, 465, 76

\bibitem[\protect\citeauthoryear{Moore, Ghigna, Governato, Lake, Quinn, Stadel  \& Tozzi}{Moore et~al.}{1999}]{moore_dark_1999}
Moore B.,  Ghigna S.,  Governato F.,  Lake G.,  Quinn T.,  Stadel J.,   Tozzi P.,  1999, \mn@doi [The Astrophysical Journal] {10.1086/312287}, 524, L19

\bibitem[\protect\citeauthoryear{Mosbech, Boehm  \& Wong}{Mosbech et~al.}{2023}]{mosbech_probing_2023}
Mosbech M.~R.,  Boehm C.,   Wong Y. Y.~Y.,  2023, \mn@doi [Journal of Cosmology and Astroparticle Physics] {10.1088/1475-7516/2023/03/047}, 2023, 047

\bibitem[\protect\citeauthoryear{Ménard, Scranton, Fukugita  \& Richards}{Ménard et~al.}{2010}]{menard_measuring_2010}
Ménard B.,  Scranton R.,  Fukugita M.,   Richards G.,  2010, \mn@doi [Monthly Notices of the Royal Astronomical Society] {10.1111/j.1365-2966.2010.16486.x}, 405, 1025

\bibitem[\protect\citeauthoryear{Nadler et~al.,}{Nadler et~al.}{2021a}]{nadler_constraints_2021}
Nadler E.~O.,  et~al., 2021a, \mn@doi [Physical Review Letters] {10.1103/PhysRevLett.126.091101}, 126, 091101

\bibitem[\protect\citeauthoryear{Nadler, Birrer, Gilman, Wechsler, Du, Benson, Nierenberg  \& Treu}{Nadler et~al.}{2021b}]{nadler_dark_2021}
Nadler E.~O.,  Birrer S.,  Gilman D.,  Wechsler R.~H.,  Du X.,  Benson A.,  Nierenberg A.~M.,   Treu T.,  2021b, \mn@doi [The Astrophysical Journal] {10.3847/1538-4357/abf9a3}, 917, 7

\bibitem[\protect\citeauthoryear{Narayanan et~al.,}{Narayanan et~al.}{2021}]{narayanan_powderday_2021}
Narayanan D.,  et~al., 2021, \mn@doi [The Astrophysical Journal Supplement Series] {10.3847/1538-4365/abc487}, 252, 12

\bibitem[\protect\citeauthoryear{Navarro, Frenk  \& White}{Navarro et~al.}{1996}]{navarro_structure_1996}
Navarro J.~F.,  Frenk C.~S.,   White S. D.~M.,  1996, \mn@doi [The Astrophysical Journal] {10.1086/177173}, 462, 563

\bibitem[\protect\citeauthoryear{Newton et~al.,}{Newton et~al.}{2021}]{newton_constraints_2021}
Newton O.,  et~al., 2021, \mn@doi [Journal of Cosmology and Astroparticle Physics] {10.1088/1475-7516/2021/08/062}, 2021, 062

\bibitem[\protect\citeauthoryear{Oman, Frenk, Crain, Lovell  \& Pfeffer}{Oman et~al.}{2024}]{oman_warm_2024}
Oman K.~A.,  Frenk C.~S.,  Crain R.~A.,  Lovell M.~R.,   Pfeffer J.,  2024, \mn@doi [Monthly Notices of the Royal Astronomical Society] {10.1093/mnras/stae1772}, 533, 67

\bibitem[\protect\citeauthoryear{Pillepich et~al.,}{Pillepich et~al.}{2018}]{pillepich_simulating_2018}
Pillepich A.,  et~al., 2018, \mn@doi [Monthly Notices of the Royal Astronomical Society] {10.1093/mnras/stx2656}, 473, 4077

\bibitem[\protect\citeauthoryear{Power, Robotham, Obreschkow, Hobbs  \& Lewis}{Power et~al.}{2016}]{power_spurious_2016}
Power C.,  Robotham A. S.~G.,  Obreschkow D.,  Hobbs A.,   Lewis G.~F.,  2016, \mn@doi [Monthly Notices of the Royal Astronomical Society] {10.1093/mnras/stw1644}, 462, 474

\bibitem[\protect\citeauthoryear{Robitaille}{Robitaille}{2011}]{robitaille_hyperion_2011}
Robitaille T.~P.,  2011, \mn@doi [Astronomy and Astrophysics] {10.1051/0004-6361/201117150}, 536, A79

\bibitem[\protect\citeauthoryear{Rubin, Burstein, Ford  \& Thonnard}{Rubin et~al.}{1985}]{rubin_rotation_1985}
Rubin V.~C.,  Burstein D.,  Ford Jr. W.~K.,   Thonnard N.,  1985, \mn@doi [The Astrophysical Journal] {10.1086/162866}, 289, 81

\bibitem[\protect\citeauthoryear{Rémy-Ruyer et~al.,}{Rémy-Ruyer et~al.}{2014}]{remy-ruyer_gas--dust_2014}
Rémy-Ruyer A.,  et~al., 2014, \mn@doi [Astronomy and Astrophysics] {10.1051/0004-6361/201322803}, 563, A31

\bibitem[\protect\citeauthoryear{Sales, Wetzel  \& Fattahi}{Sales et~al.}{2022}]{sales_baryonic_2022}
Sales L.~V.,  Wetzel A.,   Fattahi A.,  2022, \mn@doi [Nature Astronomy] {10.1038/s41550-022-01689-w}, 6, 897

\bibitem[\protect\citeauthoryear{Salpeter}{Salpeter}{1955}]{salpeter_luminosity_1955}
Salpeter E.~E.,  1955, \mn@doi [The Astrophysical Journal] {10.1086/145971}, 121, 161

\bibitem[\protect\citeauthoryear{Sawala et~al.,}{Sawala et~al.}{2016}]{sawala_apostle_2016}
Sawala T.,  et~al., 2016, \mn@doi [Monthly Notices of the Royal Astronomical Society] {10.1093/mnras/stw145}, 457, 1931

\bibitem[\protect\citeauthoryear{Schaye et~al.,}{Schaye et~al.}{2023}]{schaye_flamingo_2023}
Schaye J.,  et~al., 2023, \mn@doi [Monthly Notices of the Royal Astronomical Society] {10.1093/mnras/stad2419}

\bibitem[\protect\citeauthoryear{Silva, Granato, Bressan  \& Danese}{Silva et~al.}{1998}]{silva_modeling_1998}
Silva L.,  Granato G.~L.,  Bressan A.,   Danese L.,  1998, \mn@doi [The Astrophysical Journal] {10.1086/306476}, 509, 103

\bibitem[\protect\citeauthoryear{Somerville, Gilmore, Primack  \& Domínguez}{Somerville et~al.}{2012}]{somerville_galaxy_2012}
Somerville R.~S.,  Gilmore R.~C.,  Primack J.~R.,   Domínguez A.,  2012, \mn@doi [Monthly Notices of the Royal Astronomical Society] {10.1111/j.1365-2966.2012.20490.x}, 423, 1992

\bibitem[\protect\citeauthoryear{Song et~al.,}{Song et~al.}{2024}]{song_robust_2024}
Song D.,  et~al., 2024, \mn@doi [Monthly Notices of the Royal Astronomical Society] {10.1093/mnras/stae923}, 530, 4395

\bibitem[\protect\citeauthoryear{Springel \& Hernquist}{Springel \& Hernquist}{2002}]{springel_cosmological_2002}
Springel V.,  Hernquist L.,  2002, \mn@doi [Monthly Notices of the Royal Astronomical Society] {10.1046/j.1365-8711.2002.05445.x}, 333, 649

\bibitem[\protect\citeauthoryear{Springel \& Hernquist}{Springel \& Hernquist}{2003}]{springel_cosmological_2003}
Springel V.,  Hernquist L.,  2003, \mn@doi [Monthly Notices of the Royal Astronomical Society] {10.1046/j.1365-8711.2003.06206.x}, 339, 289

\bibitem[\protect\citeauthoryear{Springel et~al.,}{Springel et~al.}{2005}]{springel_simulating_2005}
Springel V.,  et~al., 2005, \mn@doi [Nature] {10.1038/nature03597}, 435, 629

\bibitem[\protect\citeauthoryear{Springel, Pakmor, Zier  \& Reinecke}{Springel et~al.}{2021}]{springel_simulating_2021}
Springel V.,  Pakmor R.,  Zier O.,   Reinecke M.,  2021, \mn@doi [Monthly Notices of the Royal Astronomical Society] {10.1093/mnras/stab1855}, 506, 2871

\bibitem[\protect\citeauthoryear{Strigari, Bullock, Kaplinghat, Simon, Geha, Willman  \& Walker}{Strigari et~al.}{2008}]{strigari_common_2008}
Strigari L.~E.,  Bullock J.~S.,  Kaplinghat M.,  Simon J.~D.,  Geha M.,  Willman B.,   Walker M.~G.,  2008, \mn@doi [Nature] {10.1038/nature07222}, 454, 1096

\bibitem[\protect\citeauthoryear{Todini \& Ferrara}{Todini \& Ferrara}{2001}]{todini_dust_2001}
Todini P.,  Ferrara A.,  2001, \mn@doi [Monthly Notices of the Royal Astronomical Society] {10.1046/j.1365-8711.2001.04486.x}, 325, 726

\bibitem[\protect\citeauthoryear{Turk, Smith, Oishi, Skory, Skillman, Abel  \& Norman}{Turk et~al.}{2011}]{turk_yt_2011}
Turk M.~J.,  Smith B.~D.,  Oishi J.~S.,  Skory S.,  Skillman S.~W.,  Abel T.,   Norman M.~L.,  2011, \mn@doi [The Astrophysical Journal Supplement Series] {10.1088/0067-0049/192/1/9}, 192, 9

\bibitem[\protect\citeauthoryear{Viel, Lesgourgues, Haehnelt, Matarrese  \& Riotto}{Viel et~al.}{2005}]{viel_constraining_2005}
Viel M.,  Lesgourgues J.,  Haehnelt M.~G.,  Matarrese S.,   Riotto A.,  2005, \mn@doi [Physical Review D] {10.1103/PhysRevD.71.063534}, 71, 063534

\bibitem[\protect\citeauthoryear{Viel, Becker, Bolton  \& Haehnelt}{Viel et~al.}{2013}]{viel_warm_2013}
Viel M.,  Becker G.~D.,  Bolton J.~S.,   Haehnelt M.~G.,  2013, \mn@doi [Physical Review D] {10.1103/PhysRevD.88.043502}, 88, 043502

\bibitem[\protect\citeauthoryear{Vogelsberger, McKinnon, O'Neil, Marinacci, Torrey  \& Kannan}{Vogelsberger et~al.}{2019}]{vogelsberger_dust_2019}
Vogelsberger M.,  McKinnon R.,  O'Neil S.,  Marinacci F.,  Torrey P.,   Kannan R.,  2019, \mn@doi [Monthly Notices of the Royal Astronomical Society] {10.1093/mnras/stz1644}, 487, 4870

\bibitem[\protect\citeauthoryear{Vogelsberger, Marinacci, Torrey  \& Puchwein}{Vogelsberger et~al.}{2020}]{vogelsberger_cosmological_2020}
Vogelsberger M.,  Marinacci F.,  Torrey P.,   Puchwein E.,  2020, \mn@doi [Nature Reviews Physics] {10.1038/s42254-019-0127-2}, 2, 42

\bibitem[\protect\citeauthoryear{Wang \& White}{Wang \& White}{2007}]{wang_discreteness_2007}
Wang J.,  White S. D.~M.,  2007, \mn@doi [Monthly Notices of the Royal Astronomical Society] {10.1111/j.1365-2966.2007.12053.x}, 380, 93

\bibitem[\protect\citeauthoryear{Wolf, Martinez, Bullock, Kaplinghat, Geha, Muñoz, Simon  \& Avedo}{Wolf et~al.}{2010}]{wolf_accurate_2010}
Wolf J.,  Martinez G.~D.,  Bullock J.~S.,  Kaplinghat M.,  Geha M.,  Muñoz R.~R.,  Simon J.~D.,   Avedo F.~F.,  2010, \mn@doi [Monthly Notices of the Royal Astronomical Society] {10.1111/j.1365-2966.2010.16753.x}, 406, 1220

\bibitem[\protect\citeauthoryear{Woo, Courteau  \& Dekel}{Woo et~al.}{2008}]{woo_scaling_2008}
Woo J.,  Courteau S.,   Dekel A.,  2008, \mn@doi [Monthly Notices of the Royal Astronomical Society] {10.1111/j.1365-2966.2008.13770.x}, 390, 1453

\bibitem[\protect\citeauthoryear{Xu}{Xu}{1995}]{xu_new_1995}
Xu G.,  1995, \mn@doi [The Astrophysical Journal Supplement Series] {10.1086/192166}, 98, 355

\bibitem[\protect\citeauthoryear{Yajima, Shlosman, Romano-Díaz  \& Nagamine}{Yajima et~al.}{2015}]{yajima_observational_2015}
Yajima H.,  Shlosman I.,  Romano-Díaz E.,   Nagamine K.,  2015, \mn@doi [Monthly Notices of the Royal Astronomical Society] {10.1093/mnras/stv974}, 451, 418

\bibitem[\protect\citeauthoryear{Zamojski et~al.,}{Zamojski et~al.}{2007}]{zamojski_deep_2007}
Zamojski M.~A.,  et~al., 2007, \mn@doi [The Astrophysical Journal Supplement Series] {10.1086/516593}, 172, 468

\bibitem[\protect\citeauthoryear{Zwicky}{Zwicky}{1933}]{zwicky_rotverschiebung_1933}
Zwicky F.,  1933, Helvetica Physica Acta, 6, 110

\bibitem[\protect\citeauthoryear{de Blok}{de~Blok}{2010}]{de_blok_core-cusp_2010}
de Blok W. J.~G.,  2010, \mn@doi [Advances in Astronomy] {10.1155/2010/789293}, 2010, 789293

\makeatother
\end{thebibliography}

%%%%%%%%%%%%%%%%%%%%%%%%%%%%%%%%%%%%%%%%%%%%%%%%%%

%%%%%%%%%%%%%%%%% APPENDICES %%%%%%%%%%%%%%%%%%%%%

\appendix
\section{Different dust production results}\label{sec:app_dust_comp}
The different dust mass prescriptions available in \powderday{} are discussed here, though we note they don't impact the results of our analysis. In addition to the \dtm{} method presented in Section \ref{sec:dust_production}, we also ran the metallicity-dependant gas-to-dust ratio described in \cite{remy-ruyer_gas--dust_2014}, as well as the best-fit combined gas-to-dust and \dtm{} ratios from the machine learning model described in \cite{li_dust--gas_2019}. This best-fitting model produces the most dust for each simulation out of any of the tested prescriptions, while the \dtm{} ratio produces the least. For the CDM simulation with 10 per cent supernova efficiency, the best-fit model produced a total dust mass of $29.2\times10^{7}$\msun{}, the gas-to-dust ratio a total mass of $11.2\times10^{7}$\msun{}, compared with $4.6\times10^{7}$\msun{} for the \dtm{} method. 

\begin{figure*}
    \begin{subfigure}[t]{0.32\textwidth}
        \includegraphics[width=\linewidth]{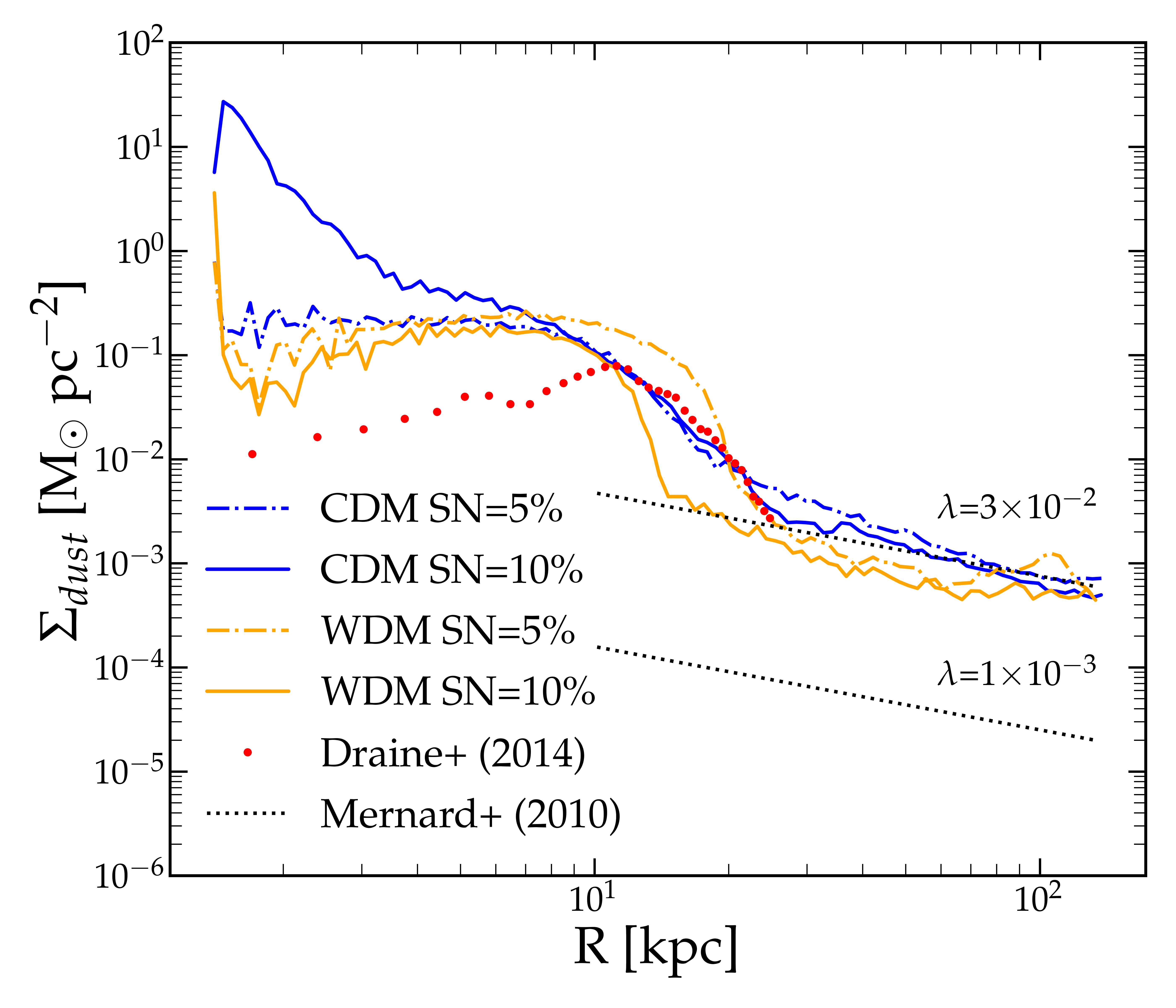}
        \caption{Dust-to-metals ratio}
        \label{fig:surf_dens_dtm}
    \end{subfigure}
    \hfill
    \begin{subfigure}[t]{0.32\textwidth}
        \includegraphics[width=\linewidth]{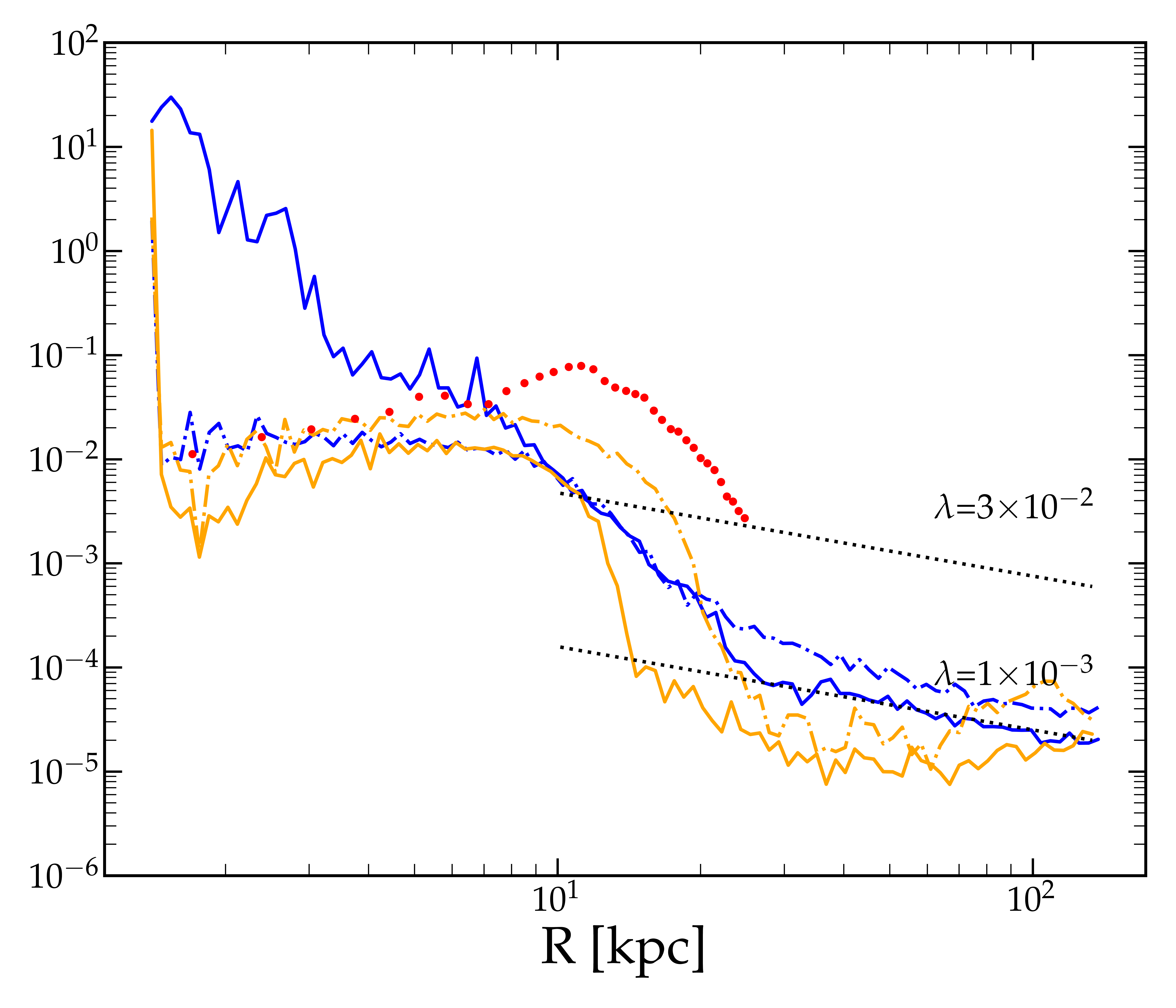}
        \caption{Remy-Ruyer dust-to-gas ratio}
        \label{fig:surf_dens_rr}
    \end{subfigure}
    \hfill
    \begin{subfigure}[t]{0.32\textwidth}
        \includegraphics[width=\linewidth]{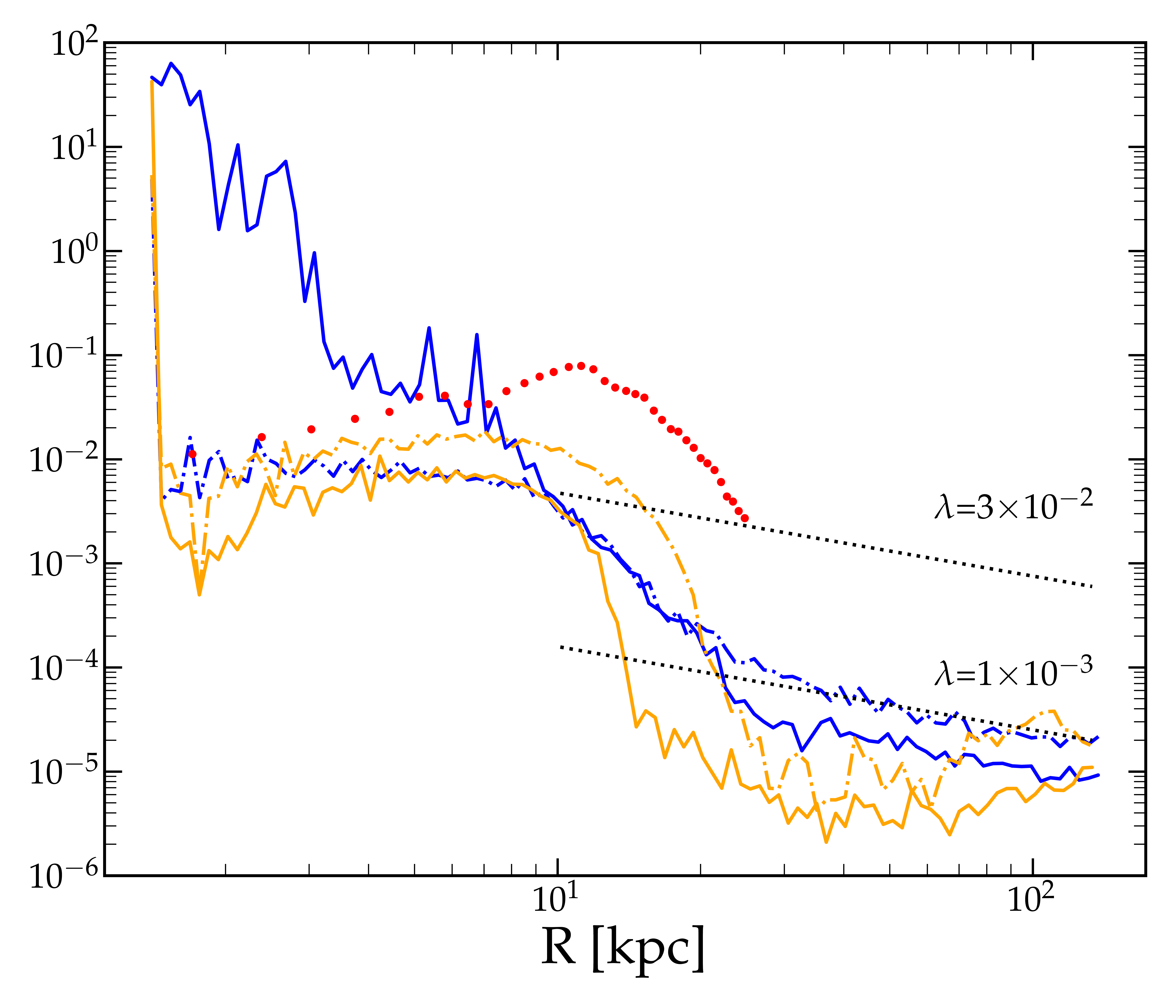}
        \caption{Li best fit models}
        \label{fig:surf_dens_li}
    \end{subfigure}
    \caption{Comparison of the dust surface density for each dust production model from \powderday{}, plus two different amplitudes for the \protect\cite{menard_measuring_2010} scaling relations, one to match the dust-to-metals ratio and the other to match both the \protect\cite{remy-ruyer_gas--dust_2014} and \protect\cite{li_dust--gas_2019} models. }
    \label{fig:all_suf_dens_mods}
\end{figure*}

We show the surface density profiles for all dust prescriptions in Fig. \ref{fig:all_suf_dens_mods}. These all have similar shapes, though slightly different amplitudes. The best-fit surface density profile, Fig. \ref{fig:surf_dens_li}, has the greatest peak amplitude in the centre of the halo but has the lowest surface density at larger radii. All dust mass prescriptions follow the \cite{menard_measuring_2010} scaling at radii $\gtrsim$15kpc, though the \dtm{} method produces comparatively higher masses in low density environments and so is fit better with a higher amplitude scaling. 

\begin{figure*}
    \begin{subfigure}[t]{0.32\textwidth}
        \includegraphics[width=\linewidth]{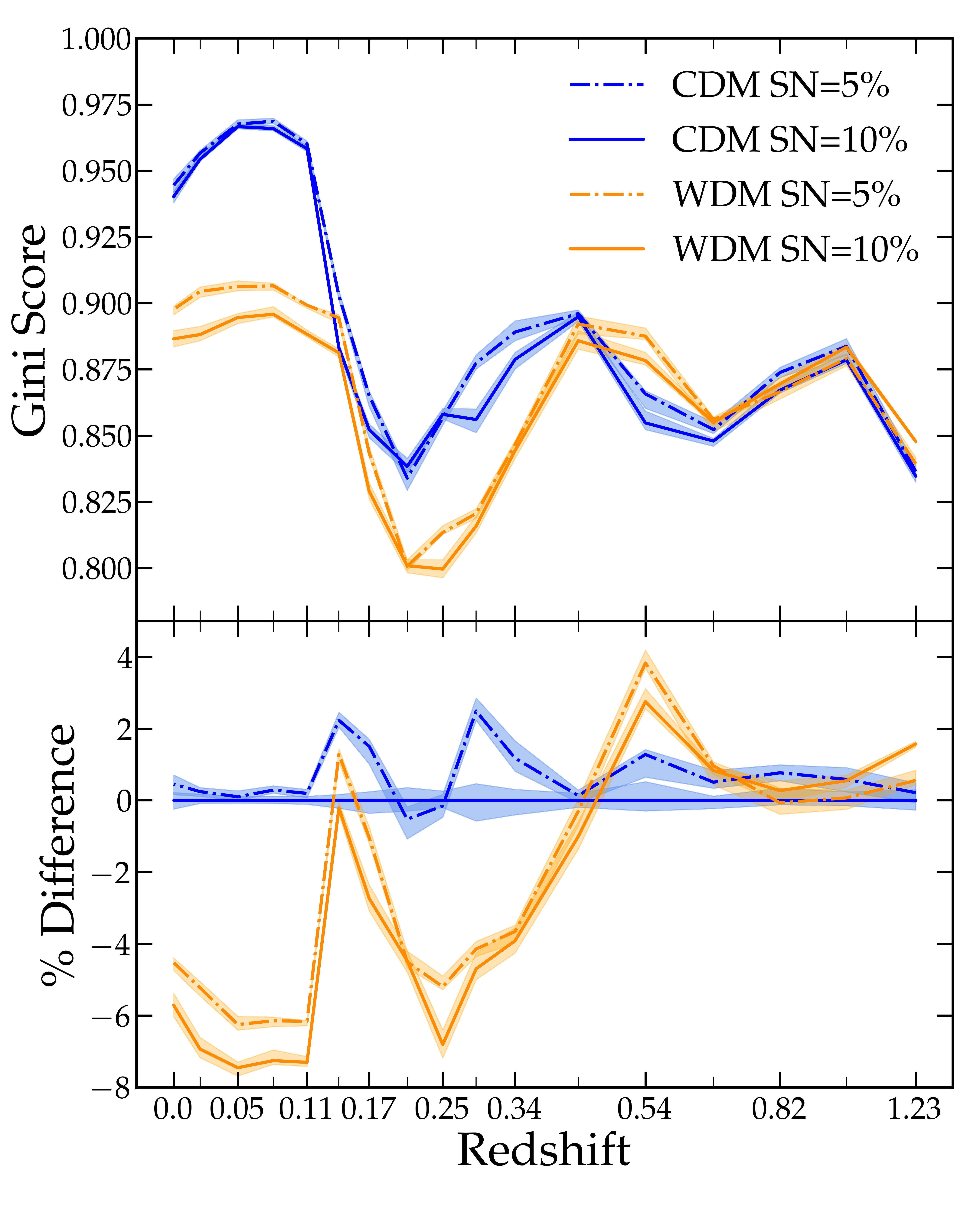}
        \caption{Dust-to-metals ratio}
        \label{fig:masked_gini_dtm}
    \end{subfigure}
    \hfill
    \begin{subfigure}[t]{0.32\textwidth}
        \includegraphics[width=\linewidth]{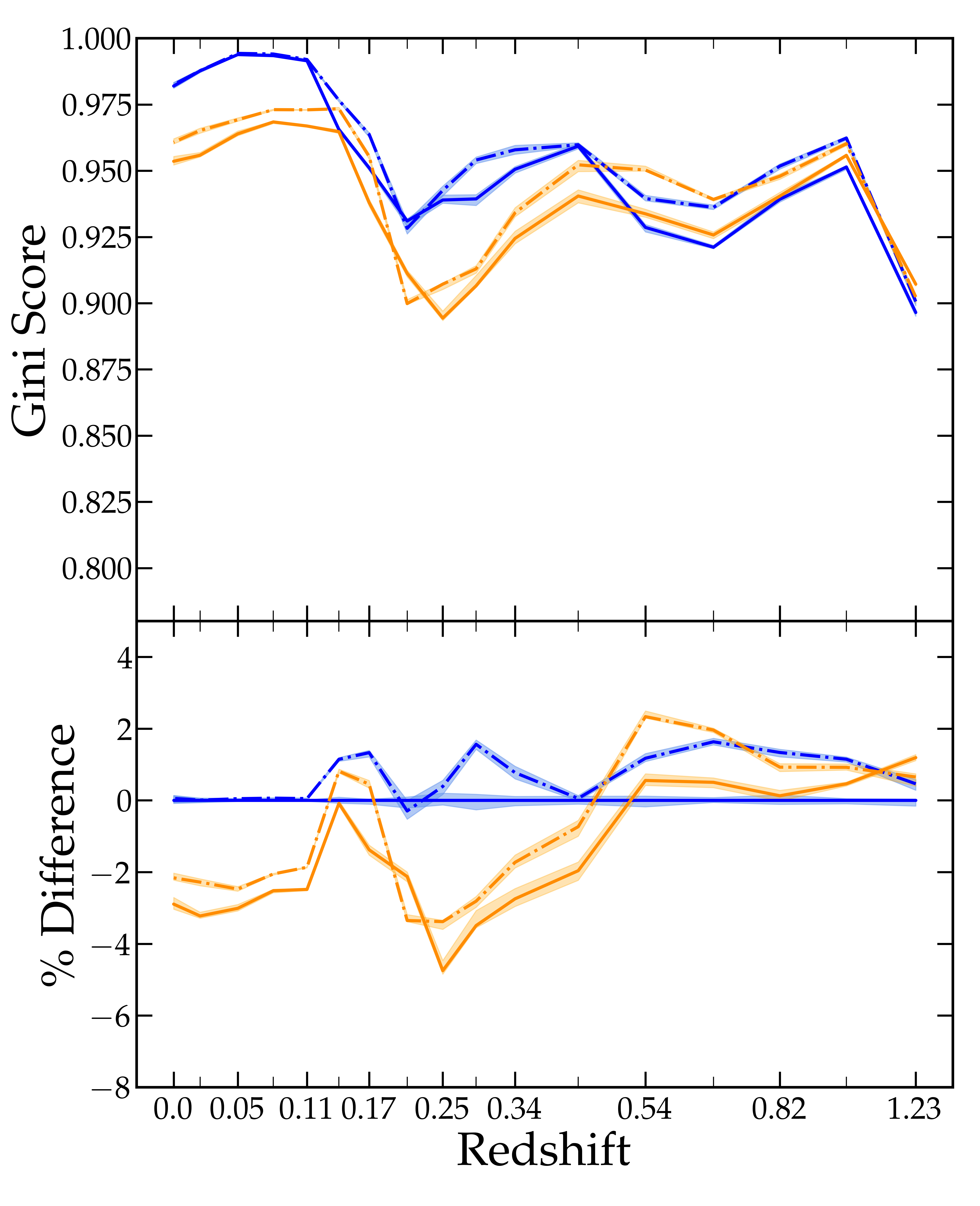}
        \caption{Remy-Ruyer gas-to-dust ratio}
        \label{fig:masked_gini_rr}
    \end{subfigure}
    \hfill
    \begin{subfigure}[t]{0.32\textwidth}
        \includegraphics[width=\linewidth]{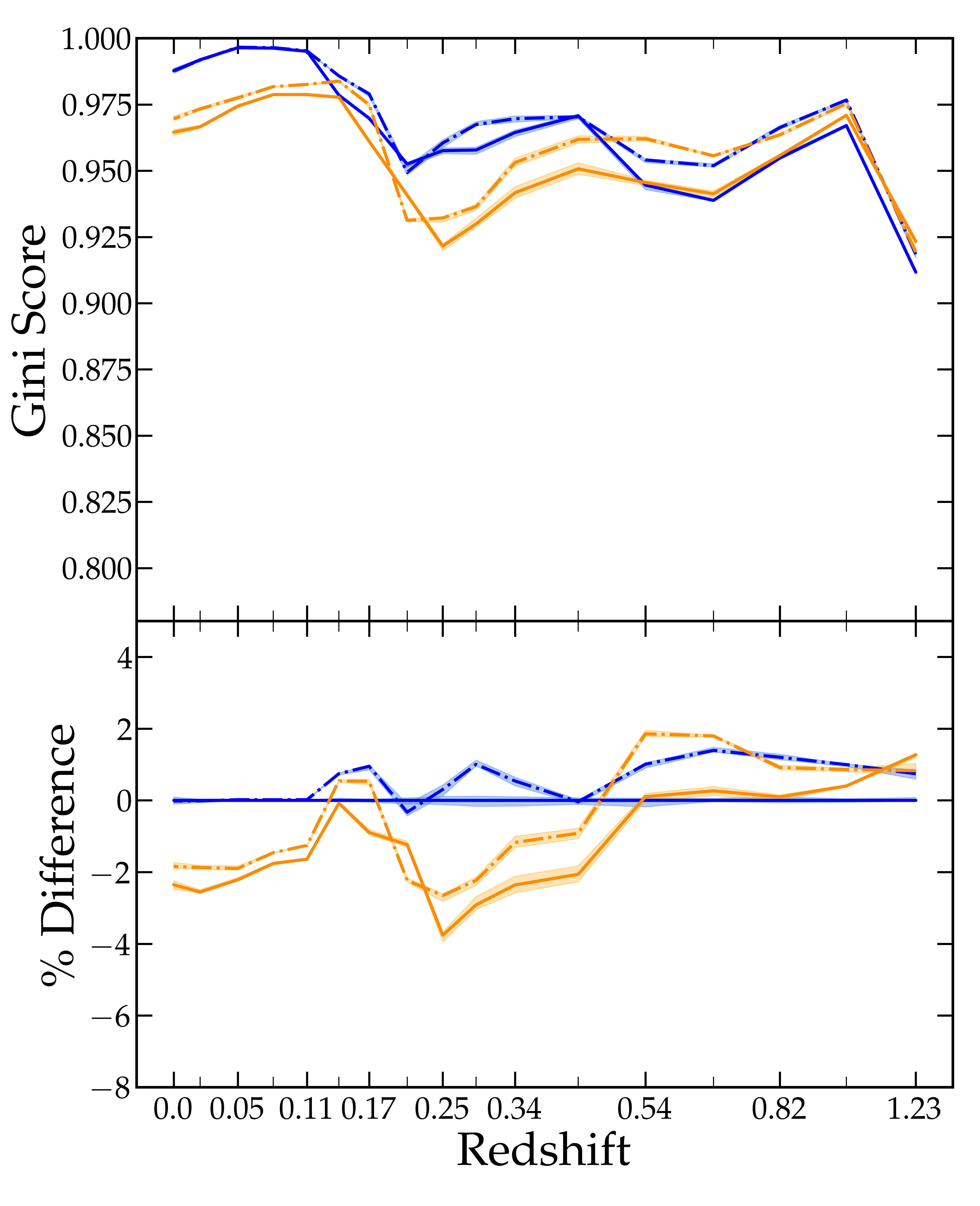}
        \caption{Li best fit models}
        \label{fig:masked_gini_li}
    \end{subfigure}
    \caption{The Gini scores (top) and percentage differences (bottom) for each dust production model from \powderday{}, plotted as a function of redshift. The amplitude of the scores differ, becoming more concentrated (higher scores) from left to right. Note, though, that the shape of the evolutions remains constant.}
    \label{fig:all_ginis}
\end{figure*}

More extreme changes in surface density are also evident in the Gini comparisons shown in Fig. \ref{fig:all_ginis}. The \dtm{} method has the smoothest transition from the inner to outer halo in surface density (Fig. \ref{fig:surf_dens_dtm}) and the least concentrated Gini evolution (Fig. \ref{fig:masked_gini_dtm}). Conversely, the best-fit model has a more extreme transition in its density profile (Fig. \ref{fig:surf_dens_li}) and a more concentrated Gini evolution (Fig. \ref{fig:masked_gini_li}). The gas-to-dust dust mass calculation method is quite similar to the best-fit model with slightly lower concentrations of dust. Despite the differences in total dust mass and surface density profiles, the difference between warm and cold dark is persistent in all models, confirming our results above.

\section{Halo masking}\label{sec:app_halo_mask}
\begin{figure*}
    \begin{subfigure}[t]{0.32\textwidth}
        \includegraphics[width=\linewidth]{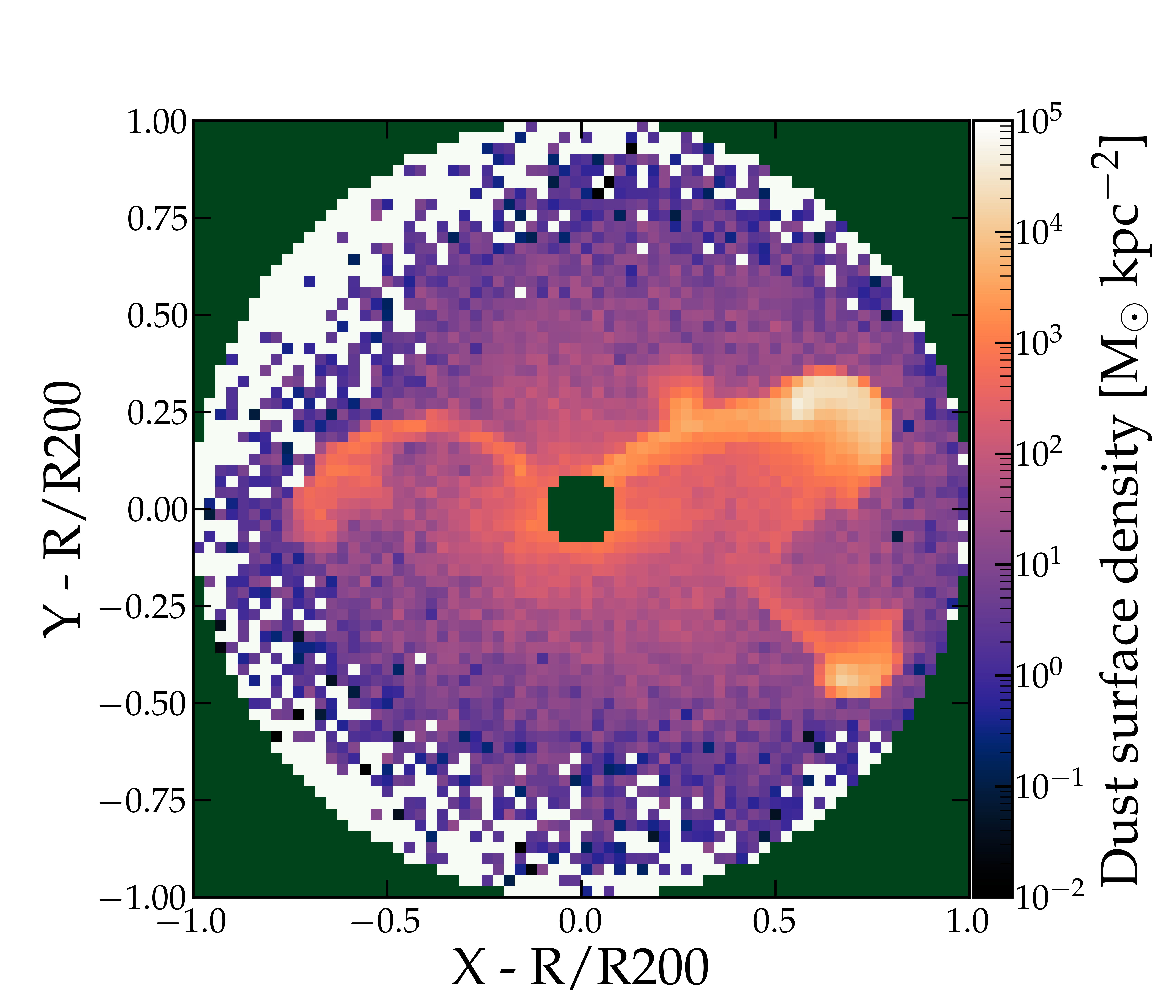}
        \caption{X-Y}
        \label{fig:masked_hists_xy}
    \end{subfigure}
    \hfill
    \begin{subfigure}[t]{0.32\textwidth}
        \includegraphics[width=\linewidth]{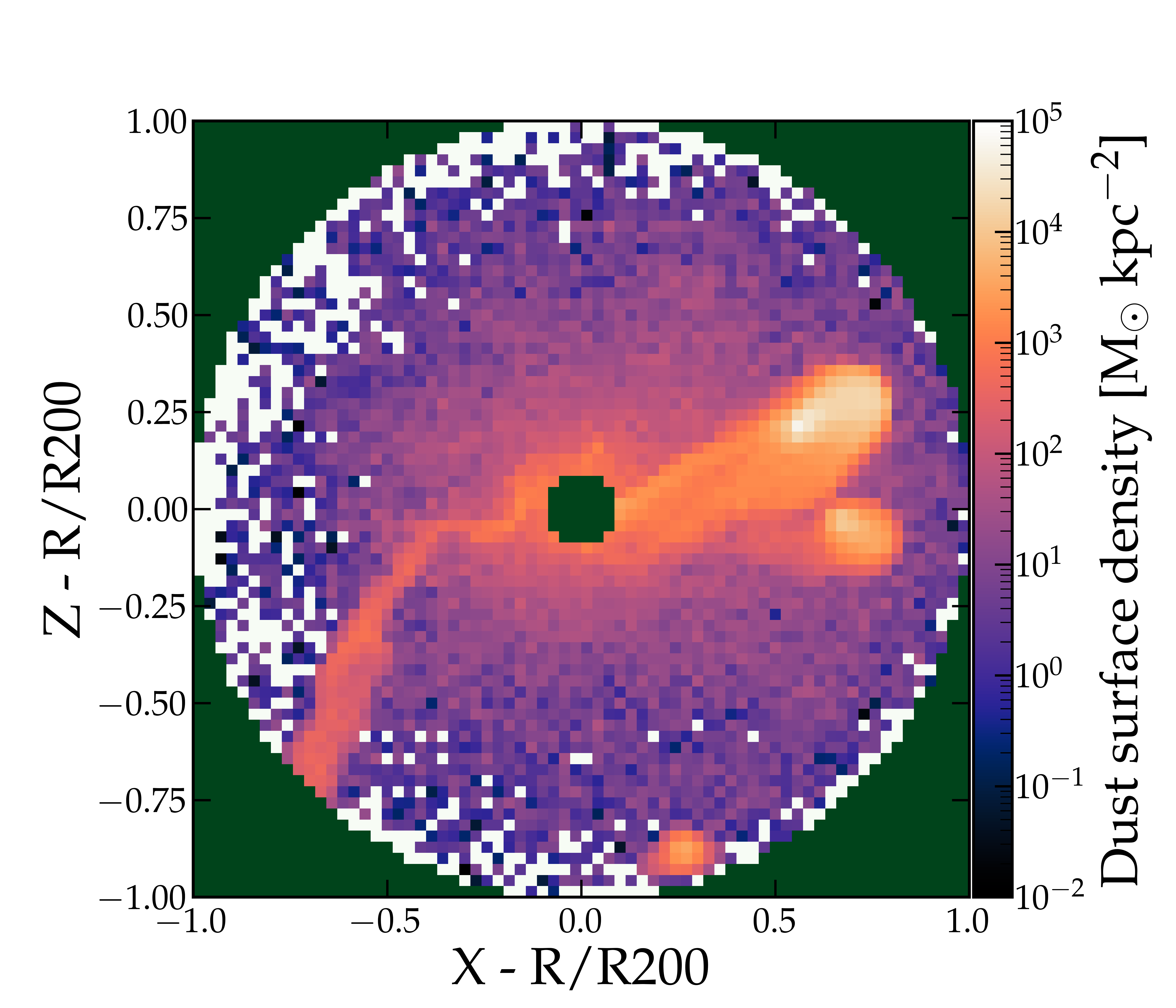}
        \caption{X-Z}
        \label{fig:masked_hists_xz}
    \end{subfigure}
    \hfill
    \begin{subfigure}[t]{0.32\textwidth}
        \includegraphics[width=\linewidth]{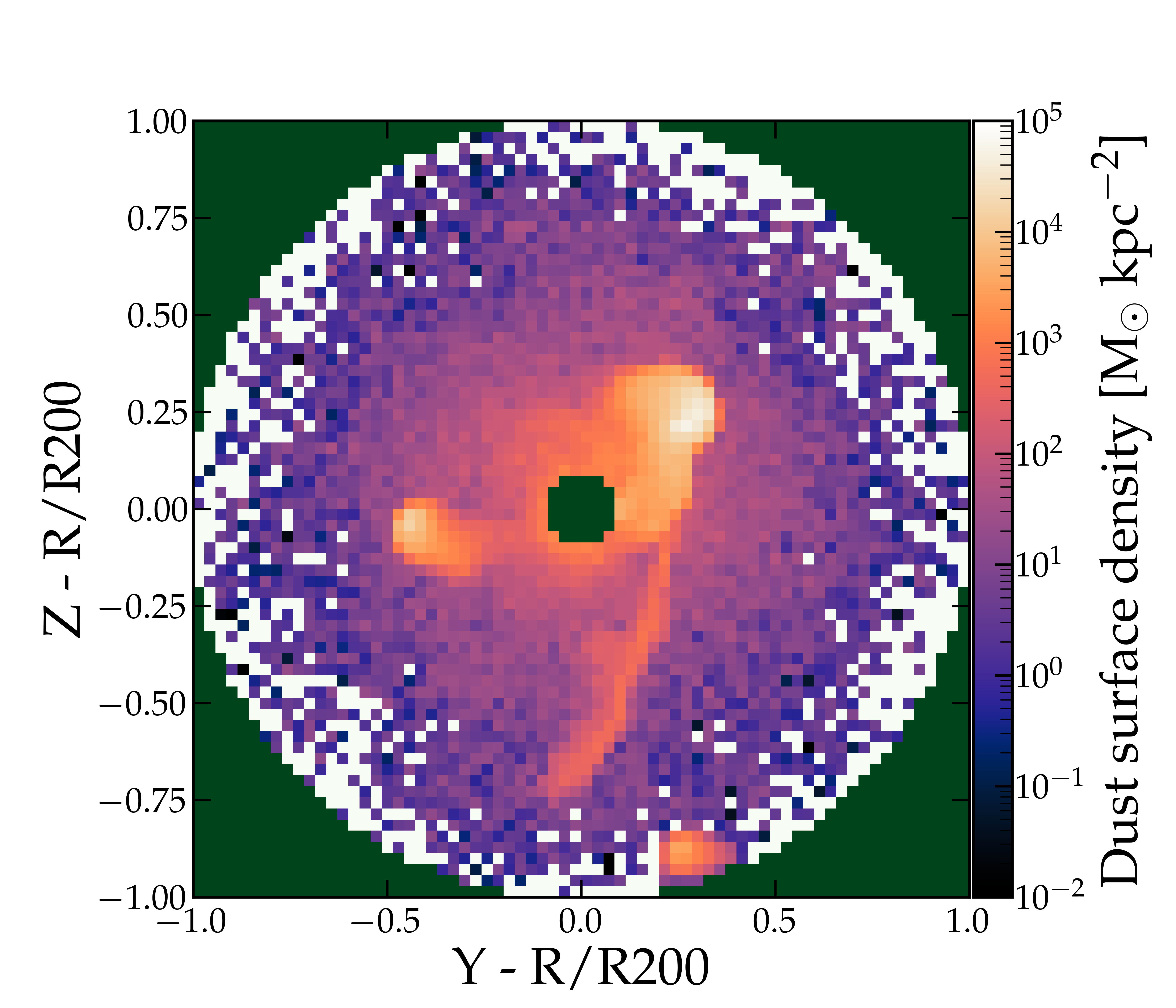}
        \caption{Y-Z}
        \label{fig:masked_hists_yz}
    \end{subfigure}
    \caption{The masking method used for the Gini calculations. Masked bins are shown in dark green, and white bins are empty. We masked everything outside of R$_{200c}$, as well as the bins with a centre within 15kpc, which would otherwise saturate any signal.}
    \label{fig:masked_hists}
\end{figure*}

Here, we show the masking method used to calculate the Gini score outlined in Section \ref{sec:gini_intro}. After creating the 2D projected histograms of dust mass similar to Fig. \ref{fig:dust_hists_all}, we take the centre of each bin, and if the centre of the bin is less than 15kpc from the centre of the halo, the bin is excluded from the calculation. Bins are also excluded if the bin centre is beyond R$_{200c}$. The inner 15kpc and R$_{200}$ limits are in comoving units for the evolution analysis. Fig. \ref{fig:masked_hists} shows these masked 2D projections of the halo, with the masked regions in dark green. The same masking is applied to all rotations we use to calculate the median Gini score and errors. 

%%%%%%%%%%%%%%%%%%%%%%%%%%%%%%%%%%%%%%%%%%%%%%%%%%

% Don't change these lines
\bsp	% typesetting comment
\label{lastpage}
\end{document}